\documentclass{article}
\usepackage{arxiv}
\usepackage{pdfpages}
\usepackage{graphicx}
\usepackage{cite}
\usepackage{verbatim}
\usepackage{color}
\usepackage{subfigure}
\usepackage{dcolumn}
\usepackage{bm}
\usepackage{floatrow}
\usepackage{amsmath}
\usepackage{amssymb}
\usepackage{makeidx}
\usepackage{epsfig}
\usepackage{amsfonts}
\usepackage{hyperref}
\usepackage{array}
\usepackage{epstopdf}
\usepackage{multirow}
\usepackage{framed}
\usepackage{mathtools}
\def\qe{\textsc{Quantum} ESPRESSO}

\newcommand{\RNum}[1]
{\uppercase\expandafter{\romannumeral #1\relax}}
\newcommand{\Rnum}[1]
{\lowercase\expandafter{\romannumeral #1\relax}}
\usepackage{color}
\usepackage{subfigure}
\usepackage{graphicx}
\usepackage{dcolumn}
\usepackage{bm}
\usepackage{braket}
\usepackage{floatrow}
\usepackage{amsmath}
\usepackage{makeidx}
\usepackage{epsfig}
\usepackage{array}
\usepackage{epstopdf}
\usepackage{multirow}
\usepackage[utf8]{inputenc} 
\usepackage[T1]{fontenc}    
\usepackage{hyperref}       
\usepackage{url}            
\usepackage{booktabs}       
\usepackage{amsfonts}       
\usepackage{nicefrac}       
\usepackage{microtype}      
\usepackage{lipsum}	
\usepackage{listofsymbols}
\usepackage{moreverb}
\usepackage{listings} 
\title{Biocompatibility of 2D silicon nitride: Interaction at the nano--bio interface}
\author{
  Ashkan Shekaari \\
  Department of Physics\\
  K. N. Toosi University of Technology\\
  Tehran, 15875-4416, Iran \\
  \texttt{shekaari@email.kntu.ac.ir} \\
   \And
 Mahmoud Jafari \thanks{Corresponding author} \\
  Department of Physics\\
  K. N. Toosi University of Technology\\
  Tehran, 15875-4416, Iran \\
  \texttt{jafari@kntu.ac.ir} \\}
\begin{document}
\maketitle
\begin{abstract}
Determining potential abilities of nanostructures to induce toxicity to biological molecules is still a convoluted challenge in the realm of nanomedicine. Based on the unprecedented achievements of two-dimensional nanomaterials in nearly all areas of applied sciences particularly medicine, we carried out all-atom molecular dynamics simulations to assess the biologically-important, yet-unmapped issue of the biocompatibility of 2D, hexagonal $\beta$-Si$_3$N$_4$ nanosheet via investigating its possible cross interactions with both human serum albumin (HSA) and p53 tumor suppressor. Examining the conventional MD indicators in the presence and absence of the monolayer revealed that hexagonal Si$_3$N$_4$ nanosheet weakly binds to these two proteins without inducing any important, dramatic change to their secondary structures, revealing accordingly the biological compatibility of the monolayer in case it is released as therapeutics or carriers in vivo. This finding was also broadly supported by the related, time-dependent behaviors of the protein-monolayer as well as the protein-water interaction energies.
\end{abstract}
\keywords{Biocompatibility\and 2D Si$_3$N$_4$\and Nano--bio interaction\and Molecular dynamics simulation}
\section{\label{sec:1}Introduction}
The stupendous advent of nanoscience and nanotechnology has enabled human being to precisely manipulate matter at one of the deepest levels of reality, the nanoscale, with a pervasive impact on nearly all branches of physical sciences, from materials science, to electronic devices, to biotechnology, leading to the emergence of multifarious interdisciplinary fields at the interface of physics, chemistry, and biology. Among such uncharted territories is nanobiotechnology~\cite{1}, which has made a categorically great contribution to modern medicine and therefore to the dawn of the nanomedicine epoch via adding functionalities to nanomaterials and then by interfacing them with biological structures, aiming at diagnosing, preventing, and treating diseases on molecular scales as well. These nanomaterials, whether as therapeutics or carriers, inevitably interact with biological entities within human body, and one ultimate goal of such so-called nanobiosystems is then their in vivo applications~\cite{2}. To this end, they must accordingly conquer several intractable challenges, which an important of them is the biocompatibility prerequisite, in the sense that nanobiomaterials should not exhibit any toxic or injurious effect on biological systems (cells, proteins, tissues, etc.). Therefore, in vivo toxicological evaluation of nanomaterials is cardinal for exploiting them in nanomedicine.

As yet, $sp^2$ carbon nanomaterials, particularly fullerenes~\cite{3}, carbon nanotubes~\cite{4}, and graphene~\cite{5}, have excited the ardor of scientists in the area of nanomedicine because they are best suited for drug delivery~\cite{6,7,8}, sensing biological targets~\cite{9,10}, biomedical imaging~\cite{11}, and cancer treatment~\cite{12}. Novel 2D layered nanomaterials such as MoS$_2$~\cite{13}, boron nitride~\cite{14}, WS$_2$~\cite{15}, and graphite-carbon nitride~\cite{17} have also captivated many researchers in biosensing and nanomedicine because of their structural similarities to graphene as a paragon of layered nanomaterial. As a result, except silica~\cite{19,20}, minimal attention has so far been paid to silicon-based nanomaterials in terms of their potential medicinal applications. Indeed, the macroscopic state of Si$_3$N$_4$ has been proved to be biocompatible and stable in vivo, and such properties, when combined with its phenomenal mechanical features~\cite{21,22,23}, make Si$_3$N$_4$ an intriguing ceramic implant material, being truly useful in some healthcare applications, particularly in orthopedic surgery~\cite{24}. 

A recent investigation on Si$_3$N$_4$ nanostructures carried out in 2020~\cite{25} has unveiled a new, 2D member of this family, showing the yet-continuing promisingness of silicon nitride materials in the post-graphene age. In the same work, it has been proved that $\beta$-Si$_3$N$_4$ nanosheets exhibit a semiconducting behavior with a band gap of about 2 eV. Taking into account the findings that semiconducting MoS$_2$~\cite{26} or graphene~\cite{27} monolayers indeed destabilize amyloid beta fibrils~\cite{28}, we therefore hypothesized that 2D Si$_3$N$_4$ may exhibit the same effect based on the structural and electronic similarities. As a result, the present work was devoted to evaluating the biocompatibility of $\beta$-Si$_3$N$_4$ monolayers as a first step in testing our hypothesis in case they are released in vivo. We carried out the present investigation via examining possible cross interactions of the monolayer with two specific homo sapiens proteins, namely human serum albumin (HSA)---as the most abundant protein in human blood plasma, constituting about half of the serum protein---and the p53 tumor suppressor protein described as the guardian of the genome due to its role in conserving genome stability by preventing mutations~\cite{29}. We applied all-atom molecular dynamics (MD) simulations, and calculated and examined a number of important conventional MD indicators as described in Sec.~\ref{sec:2}.
\section{\label{sec:2}Computational details}
\begin{figure}[H]
	\centering
	\fbox{\rule[0cm]{0cm}{0cm}     \rule[0cm]{0cm}{0cm}
	\subfigure[]{\label{subfig:1(a)}
		\includegraphics[scale=0.047]{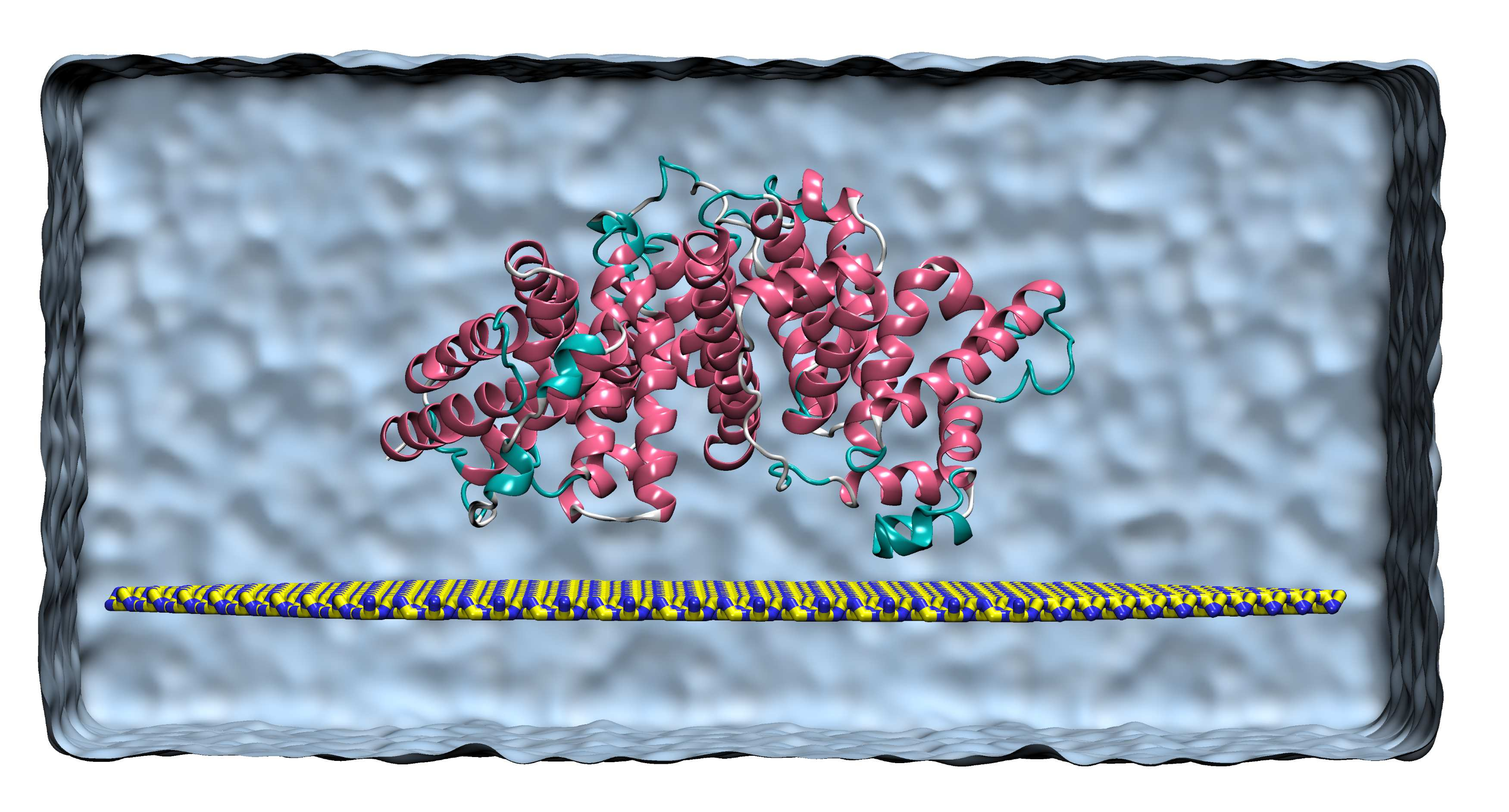}}
	\subfigure[]{\label{subfig:1(b)}
		\includegraphics[scale=0.055]{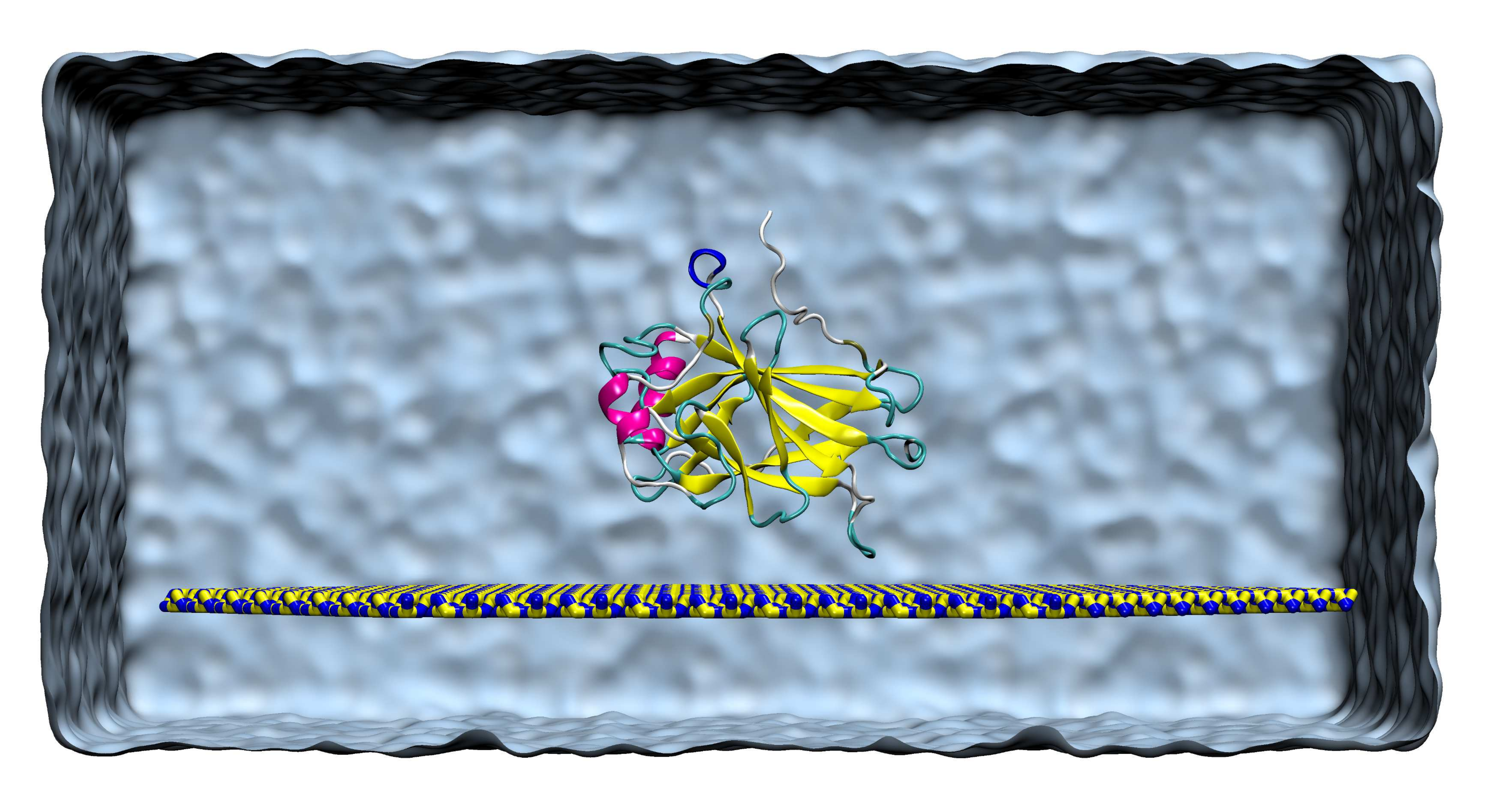}}}
	\caption{\label{fig:1}
		The simulated nanobiosystems including (a) HSA-ML-W, and (b) p53-ML-W---rendered in VMD using Tachyon parallel$\big/$multiprocessor ray tracing system~\cite{39}. The $\alpha$-helices, $\beta$-strands, and random coils$\big/$turns have been shown in pink, yellow, and cyan$\big/$white, respectively.}
\end{figure}

Initial atomic positions of HSA and p53 proteins were taken from the RCSB Protein Data Bank (PDB)~\cite{30} with the entry codes 3B9M and 1TUP, respectively. All the classical MD simulations were carried out by NAMD (version 2.14b2)~\cite{31}, in parallel, on Debian-style~\cite{32} Linux~\cite{33} systems using Open MPI v.3.1.6~\cite{34}, with the July 2018 update of CHARMM36~\cite{35,36} force fields. The VMD program (version 1.9.4a9)~\cite{37} was also used for post-processing. The two nanobiosystems including HSA-monolayer and p53-monolayer (abbreviated as HSA-ML-W and p53-ML-W; ML for monolayer and W for water) were then solvated in two boxes with TIP3P~\cite{38} water with dimensions of $17.8\times 15.5\times 8.7$, and $17.8\times 15.5\times 7.5$ nm$^3$ respectively, under periodic boundary conditions with a unit-cell padding of about 2 nm to decouple the periodic interactions, as illustrated in Fig.~\ref{fig:1}. 
We also solvated the two proteins in the same water boxes (abbreviated as HSA-W and p53-W), this time without Si$_3$N$_4$ monolayer, and accordingly referred to them as the reference (control) trajectories to which the results of the HSA-ML-W and p53-ML-W simulations were compared. The Na$^+$ and Cl$^-$ ions were randomly distributed (replaced by the same number of water molecules) in order for the entire systems to be electrostatically neutral, as tabulated in Table~\ref{tab:1}. 
\begin{table}[H]
	\small
	\caption{The number of ions added to the systems for making them electrostatically neutral, as well as the net charge values before and after neutralization. Each ion was replaced by one water molecule.}
	\label{tab:1}
	\begin{tabular}{l|cr|cc|c}
		\hline
		\multicolumn{1}{c|}{\raisebox{-1ex}{system}}
		& \multicolumn{2}{c|}{\raisebox{-0.8ex}{number of ions added}}
		& \multicolumn{2}{c|}{\raisebox{-0.8ex}{net charge (e)}}
		& \multicolumn{1}{c}{\raisebox{-0.8ex}{number of water}}\\
		\cline{2-3}
		\cline{4-5}
		\multicolumn{1}{c|}{\raisebox{-0.1ex}{}}
		& \multicolumn{1}{c}{\raisebox{-0.1ex}{Na$^{+}$}}
		& \multicolumn{1}{r|}{\raisebox{-0.1ex}{Cl$^{-}$}}
		& \multicolumn{1}{c}{\raisebox{-0.1ex}{before}}
		& \multicolumn{1}{c|}{\raisebox{-0.1ex}{after}}
		& \multicolumn{1}{c}{\raisebox{-0.1ex}{molecules removed}}\\
		\hline
		\raisebox{-0.8ex} {HSA-W} & \raisebox{-0.8ex} {14} & \raisebox{-0.8ex} {0} & \raisebox{-0.8ex} {-14} & \raisebox{-0.8ex} {8.4$\times 10^{-6}$}&14\\
		\raisebox{-0.8ex} {HSA-ML-W} & \raisebox{-0.8ex} {14} & \raisebox{-0.8ex} {0}  & \raisebox{-0.8ex} {-14} & \raisebox{-0.8ex} {4.6$\times 10^{-4}$}&14\\
		\raisebox{-0.8ex} {p53-W} & \raisebox{-0.8ex} {0} & \raisebox{-0.8ex} {3} & \raisebox{-0.8ex} {+3} & \raisebox{-0.8ex} {2.0$\times 10^{-6}$}&3\\
		\raisebox{-0.8ex} {p53-ML-W} & \raisebox{-0.8ex} {0} & \raisebox{-0.8ex} {3} & \raisebox{-0.8ex}{+3} & \raisebox{-0.8ex} {3.7$\times 10^{-4}$}&3\\
		\hline
	\end{tabular}
\end{table}
The switching and cutoff distances of 1.0 and 1.2 nm were also used for truncating non-bonded van der Waals interactions, respectively. Particle-mesh Ewald (PME)~\cite{40} with grid dimensions of $180\times160\times88$ and $180\times160\times80$ were applied to systems containing HSA and p53 for long-range interactions, respectively. Four minimization simulations for HSA-ML-W, p53-ML-W, HSA-W, and p53-W were carried out, each for 400000 conjugate gradient steps (0.4 ns). The next four MD simulations were carried out, each for 50 ns, with the integration time step of 1 ns within the NPT ensemble at 310 K and 1.01325 bar using Langevin forces with the Langevin damping constant of 2.5$\big/$ps along with the Nos{\'e}-Hoover Langevin piston pressure control. For systems without the silicon nitride monolayer (HSA-W and p53-W), the dielectric constant was set to 1. The hydrogen donor-acceptor distance and the angle cutoff have been respectively set to 3 \AA\ and 20$^\circ$ to identify hydrogen bonds. The solvent-accessible surface area (SASA)~\cite{41} was calculated for each system using the rolling-ball algorithm~\cite{42} with the radius of 1.4 {\AA} for the probe sphere. The interaction energy ($E_{int}=$ electrostatic$+$van der Waals energies) between protein-monolayer as well as protein-water in the presence and absence of the monolayer were also calculated as functions of time for each protein in order to check whether stable nano--bio complexes are made.
\section{\label{sec:3}Force field of $\beta$-Si$_3$N$_4$ nanosheet}
\begin{figure}[H]
	\centering
	\fbox{\rule[0cm]{0cm}{0cm}     \rule[0cm]{0cm}{0cm}
	\includegraphics[scale=0.5]{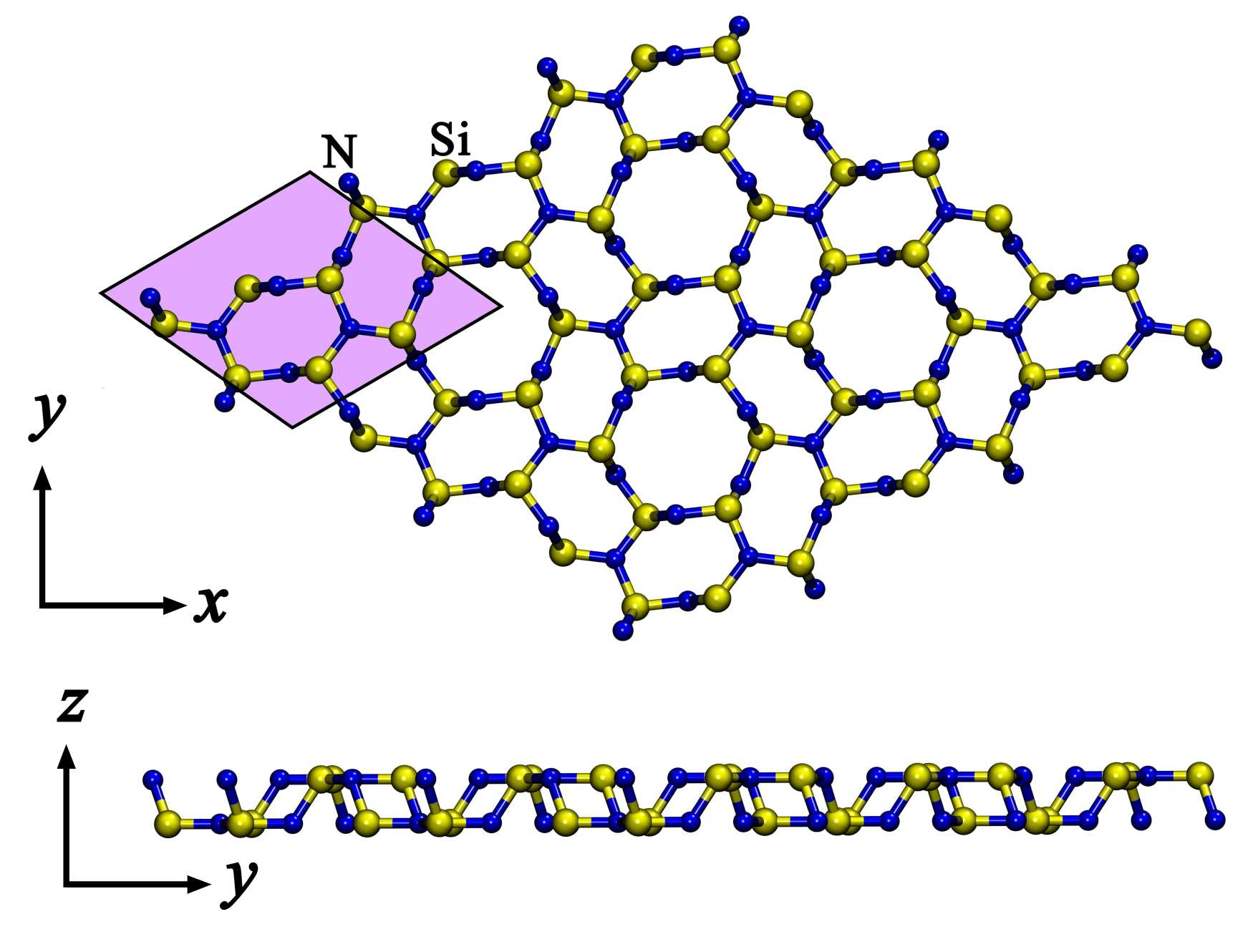}}
	\caption{\label{fig:2}
		Atomic structure of $\beta$-Si$_3$N$_4$ monolayer in the two Cartesian planes---rendered in XCrySDen~\cite{56}. The purple parallelogram is the associated unit cell, containing six Si and eight N atoms in a hexagonal Bravais lattice.}
\end{figure}

The initial force field describing the bulk phase of $\beta$-Si$_3$N$_4$ was constructed using the values provided by Wendel {\em{et al.}}~\cite{43}. The force field associated to 2D Si$_3$N$_4$ was then obtained via calibrating that of the bulk phase in a way that the (molecular-mechanical) dielectric constant of the monolayer obtained by NAMD was fitted to its (quantum-mechanical) analogue (namely, $\kappa\simeq 2.37$) calculated using \qe~\cite{44}. To this end, we adopted a self-consistent~\cite{45}, plane-wave, pseudopotential approach~\cite{46} at the PBE-GGA~\cite{47} level of density-functional theory (DFT)~\cite{48}. Scalar-relativistic ultrasoft pseudopotentials~\cite{49,50} ($\mathtt{Si.pbe}$-$\mathtt{n}$-$\mathtt{rrkjus}$\_$\mathtt{psl.1.0.0.UPF}$ and $\mathtt{N.pbe}$-$\mathtt{n}$-$\mathtt{rrkjus}$\_$\mathtt{psl.1.0.0.UPF}$~\cite{51}) generated by Rappe-Rabe-Kaxiras-Joannopoulos (RRKJ)~\cite{52} pseudization method with nonlinear core correction~\cite{53} were used to model the core electrons. The valence shells of the N and Si atomic species were also described by $2s2p$ and $3s3p$ orbitals, respectively. The pseudo-wavefunctions and charge density were expanded in a plane-wave basis set with kinetic energy cutoff values of 90 and 360 Rydberg (Ry) respectively, for which energy convergence was optimally achieved. We started from the $\beta$-Si$_3$N$_4$ primitive cell of the bulk phase containing 6 silicon and 8 nitrogen atoms in a hexagonal Bravais lattice under periodic boundary conditions with experimental lattice constant of $a_{0}=7.82$ {\AA}, and then applied a vacuum space of $>15$ \AA\ along $z$ to produce the 2D structure, as shown in Fig.~\ref{fig:2}. The equilibrium (zero-pressure) lattice constant of the nanosheet was also calculated using Murnaghan's isothermal equation of state~\cite{54,55}, leading to the theoretical, PBE-GGA value of $a=8.262$ {\AA}, which is about 5.65\% larger than that of the bulk phase ($a_{0}$).
We carried out four density-functional molecular dynamics (DFMD) simulations (i.e., two minimization and two finite-temperature free dynamics simulations) within the Car-Parrinello (CP) approach at 310 K, one with applying a zero electric field ($E=0$), and the other with $E=16$ kcal$\big/$(mol.{\AA}.e) along $+z$ to change the dipole moment ($p$) of the monolayer. These CPMD simulations were preceded by electronic minimization processes for each value of $E$ in order to bring the electronic wavefunctions on their ground states relative to the starting atomic configurations. The two setups ($E=0$ and $E=16$) were minimized after 100 damped-dynamics steps (0.012 ps) with the time-step of $\mathrm{\Delta} t=0.12$ fs and with the electron damping value ($=$ damping frequency times $\mathrm{\Delta} t$) of 0.1. The fictitious electron mass in the CP Lagrangian~\cite{57} was set to 1000 a.u. (1000 times the rest mass of electron) to guarantee the validity of the adiabatic approximation~\cite{58}; the mass cutoff of 2.5 Ry was also chosen for the Fourier acceleration effective mass to keep the quality of simulations from being adversely affected, as well as to minimize the electron drag effect. The four DFMD simulations were started from the same minimized structure in terms of ionic degrees of freedom, in that the value of the total force exerted on each atom eventually became $<10^{-4}$ eV$\big/${\AA}. The propagation time was also chosen about 0.36 ps (3000 Verlet steps). Electronic equations of motion were also accelerated using a preconditioning scheme~\cite{59}. We ignored at least the first 0.12 ps (1000 steps) of the simulations for thermalization and reliable statistical averaging. Both electronic and ionic contributions were taken into account in estimating the average value of the total dipole moment. The dielectric constant was also calculated using
\begin{equation*}
\kappa=1+\frac{\mathrm{\Delta} p}{\epsilon_{0}E\mathrm{\mathrm{\Omega}}},
\end{equation*}
where $\mathrm{\Delta} p=|p_{(E=0)}-p_{(E=16)}|$, $\epsilon_{0}=2.398\times 10^{-4}$ mol.e$^2\big/$(kcal.\AA) is the vacuum permittivity, and $\mathrm{\Omega}$ is volume of the monolayer.
\section{\label{sec:4}Results and discussion}
\begin{figure}[H]
	\centering
	\fbox{\rule[0cm]{0cm}{0cm}     \rule[0cm]{0cm}{0cm}
	\subfigure[]{\label{subfig:3(a)}
		\includegraphics[scale=0.28]{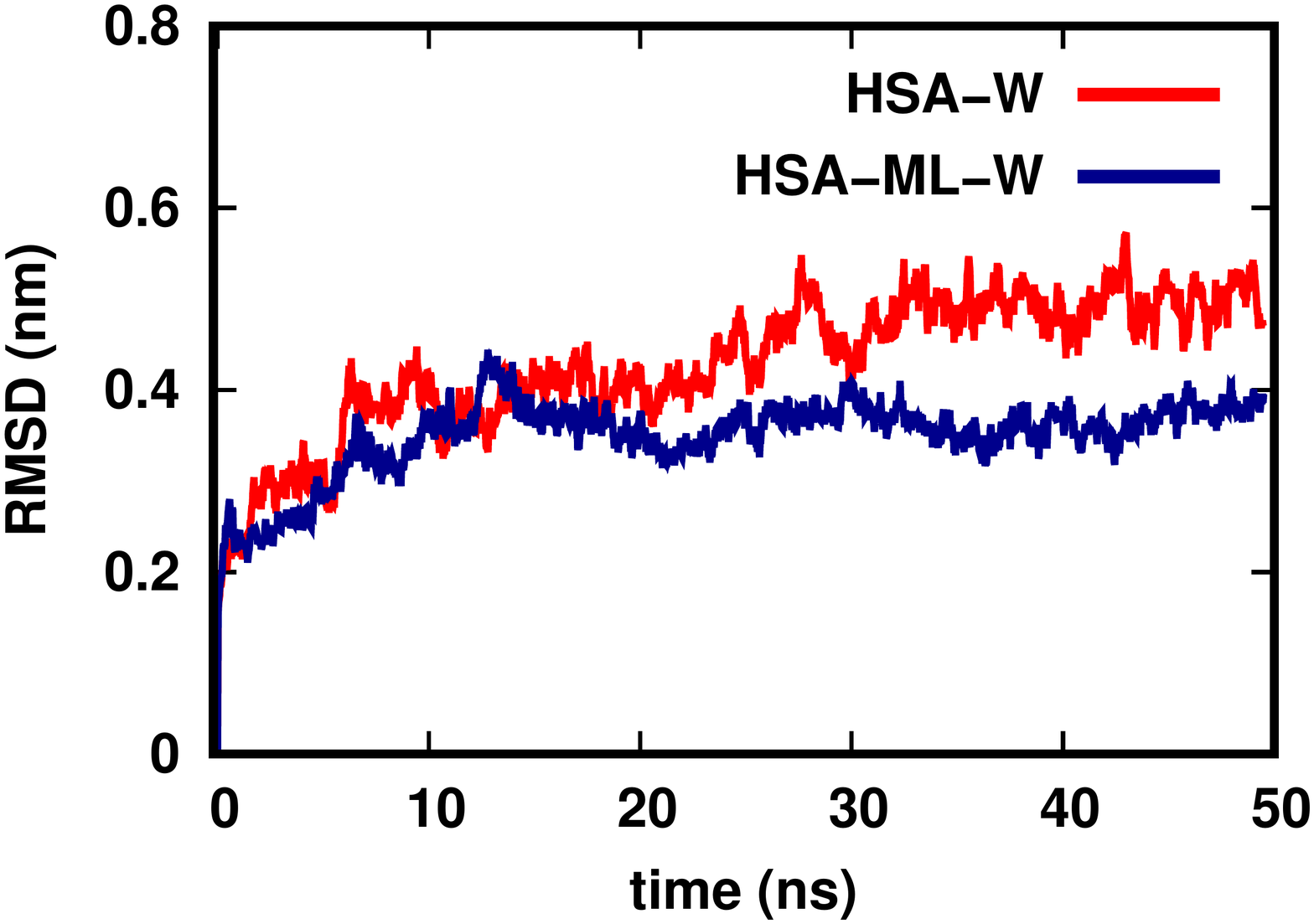}}
	\subfigure[]{\label{subfig:3(b)}
		\includegraphics[scale=0.28]{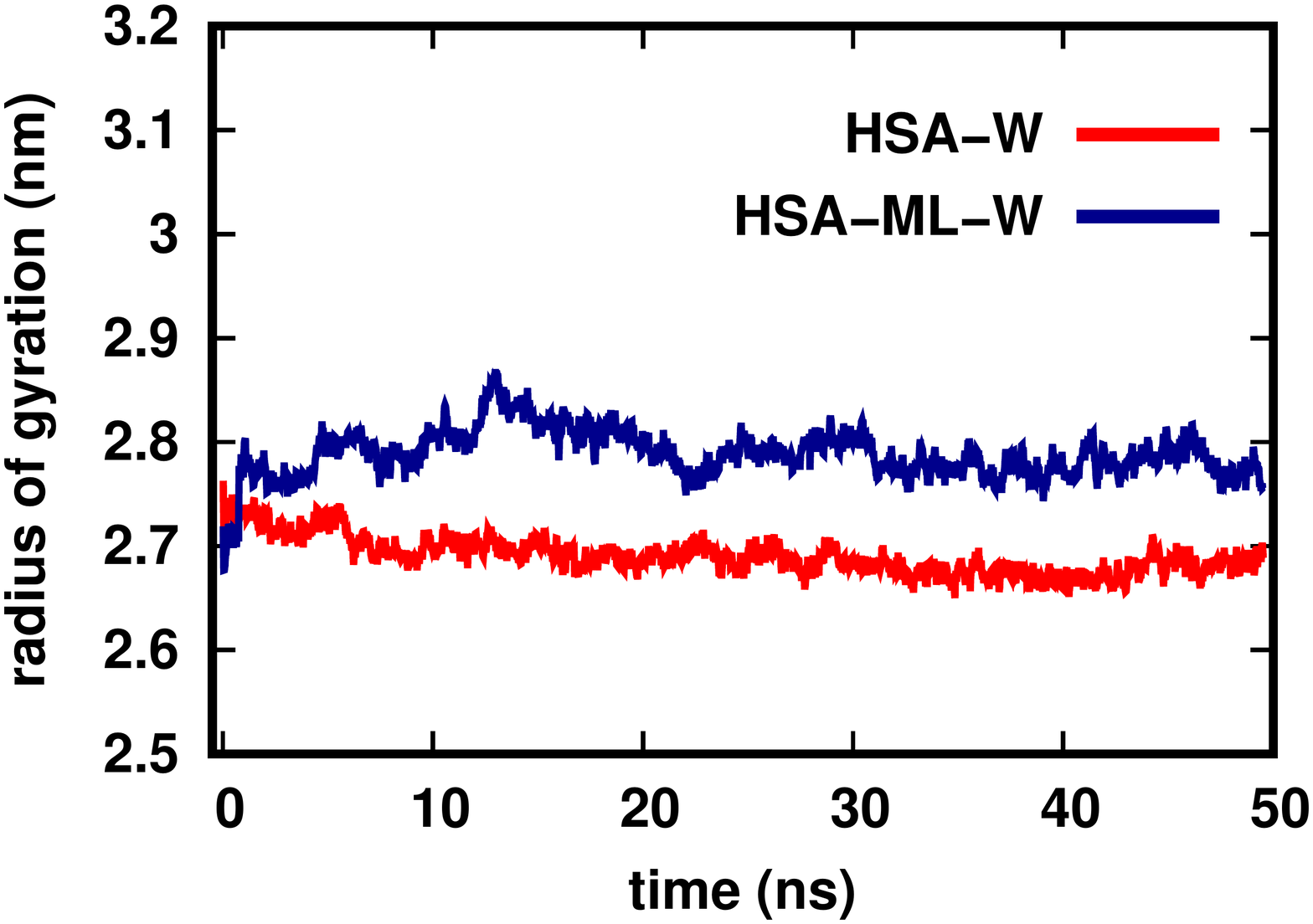}}}

\fbox{\rule[0cm]{0cm}{0cm}     \rule[0cm]{0cm}{0cm}
	\subfigure[]{\label{subfig:3(c)}
		\includegraphics[scale=0.28]{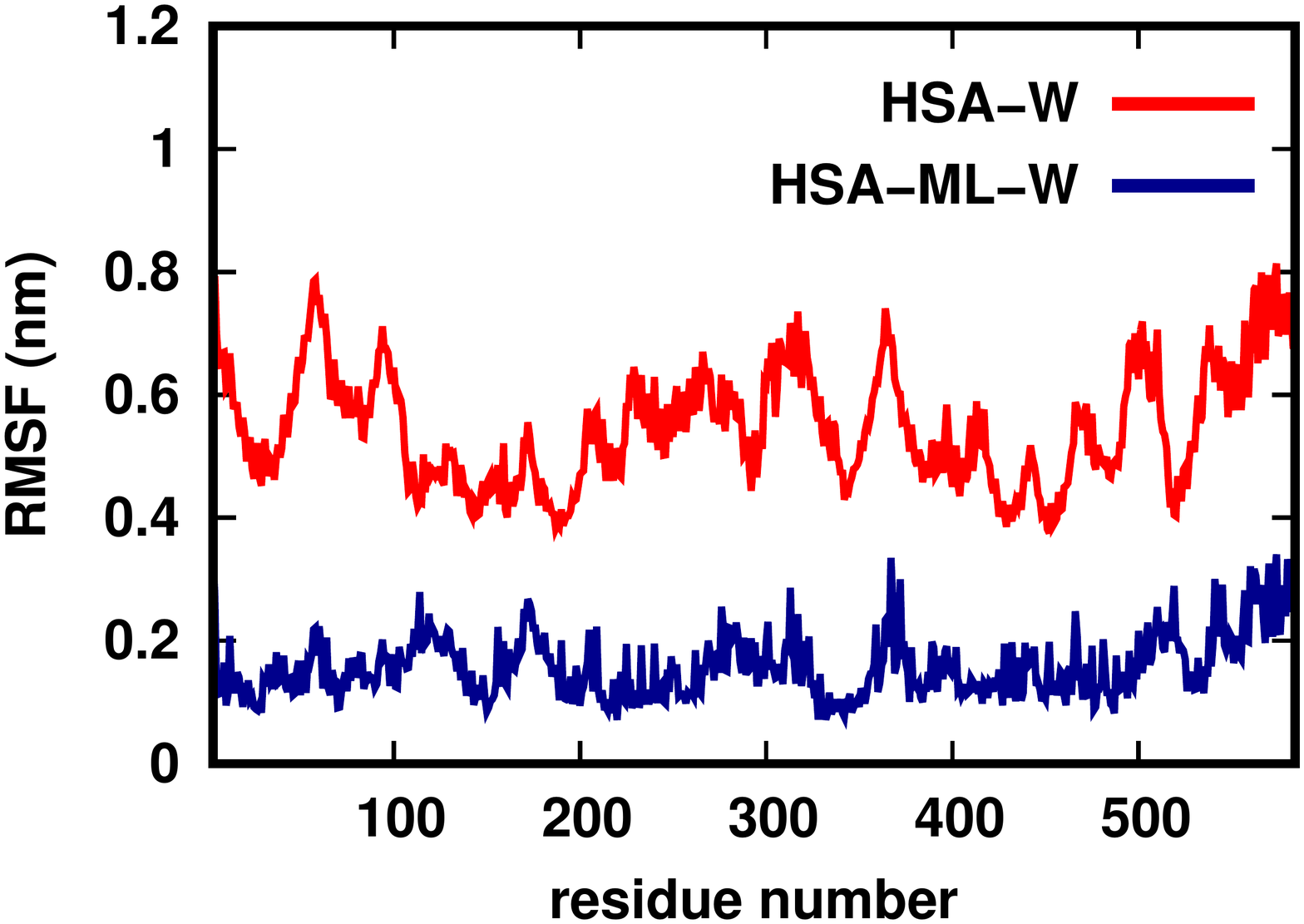}}
	\subfigure[]{\label{subfig:3(d)}
		\includegraphics[scale=0.28]{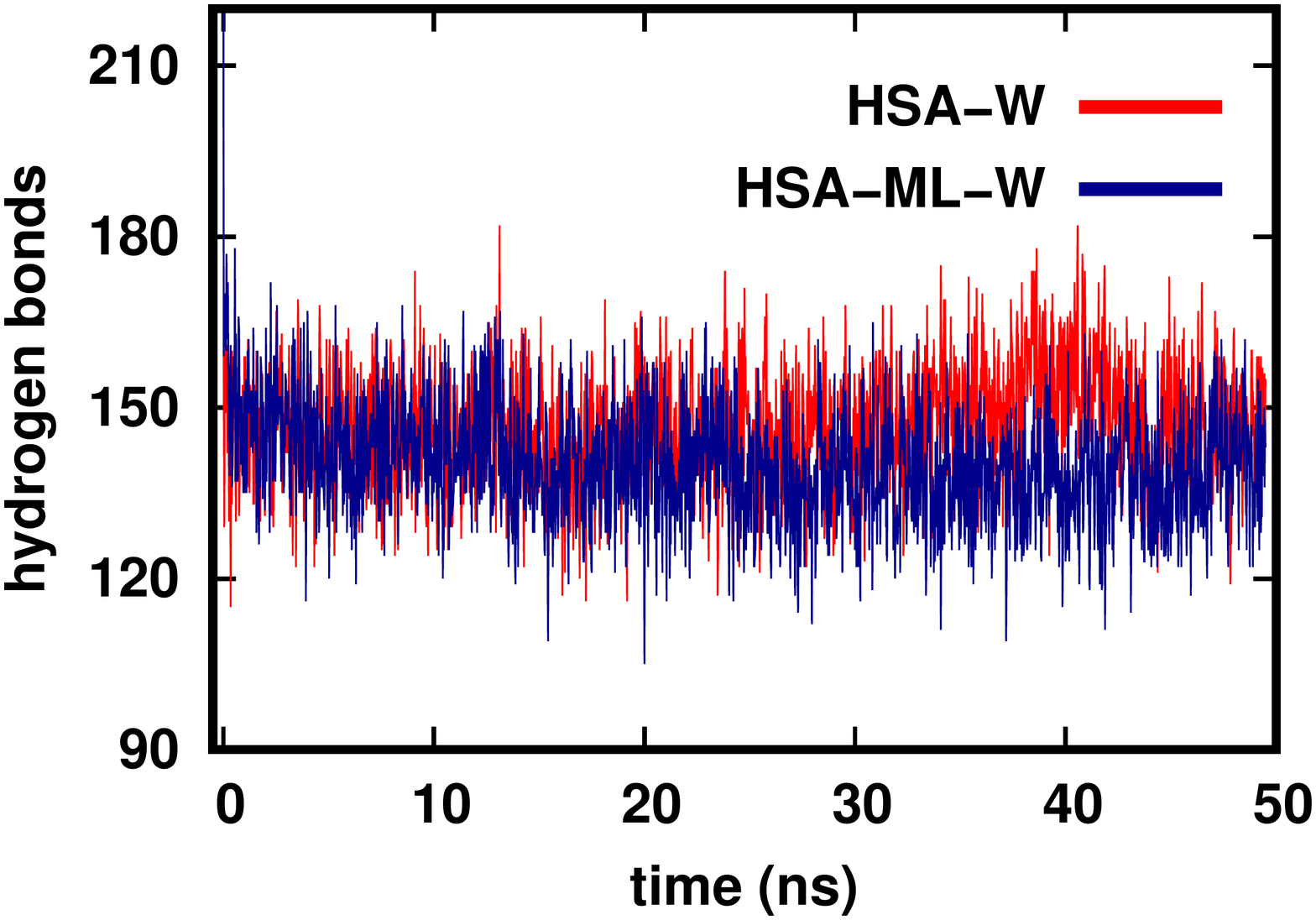}}}

\fbox{\rule[0cm]{0cm}{0cm}     \rule[0cm]{0cm}{0cm}
	\subfigure[]{\label{subfig:3(e)}
		\includegraphics[scale=0.28]{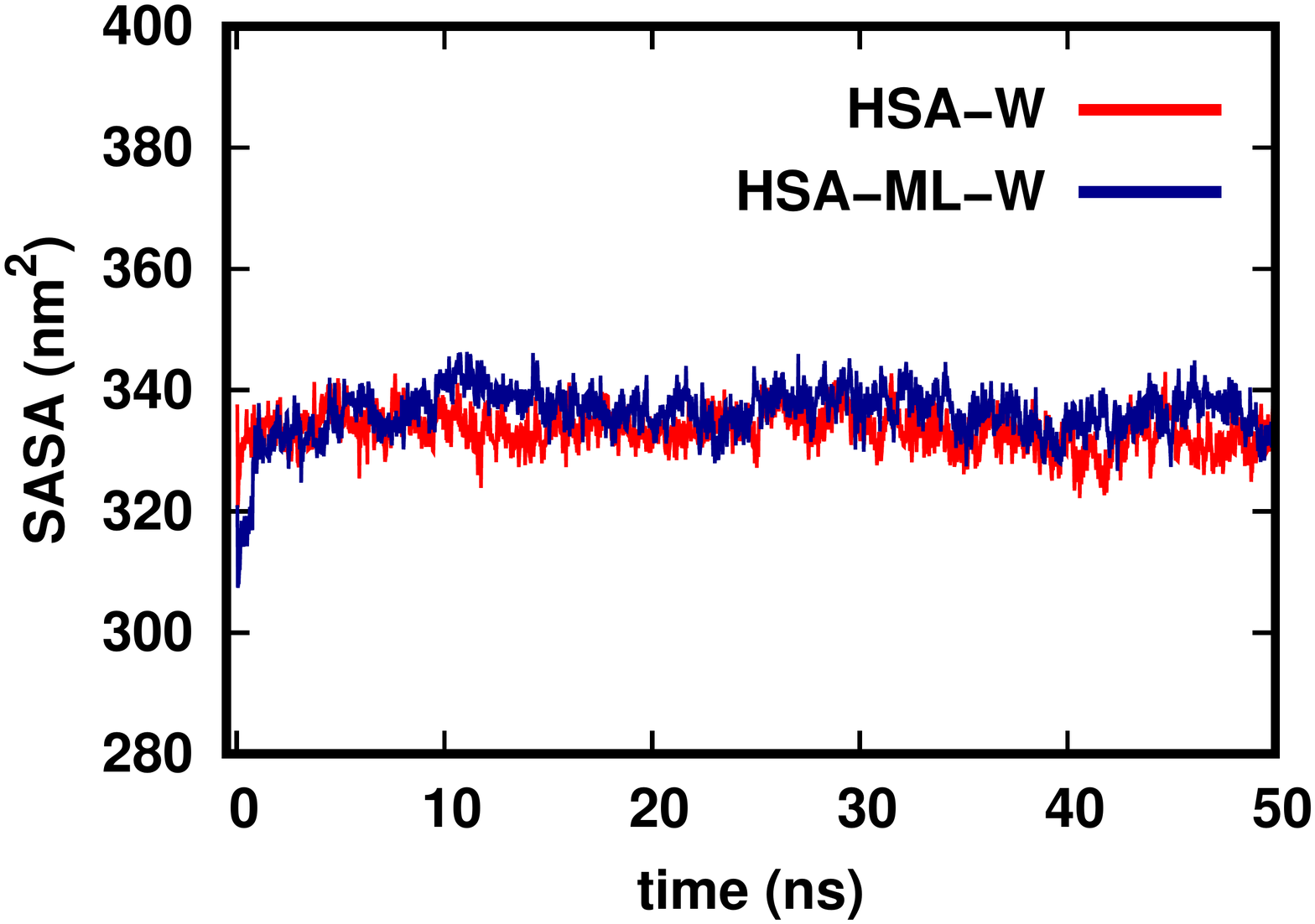}}}
	\caption{\label{fig:3}
		(a) RMSD, (b) radius of gyration, (c) per-residue RMSF, (d) number of hydrogen bonds, and (e) SASA calculated for HSA in the absence (HSA-W, red) and presence (HSA-ML-W, blue) of Si$_3$N$_4$ monolayer---rendered in Gnuplot (version 5.2)~\cite{60}.}
\end{figure}

Fig.~\ref{fig:3} illustrates the RMSD, gyration radius, per-residue RMSF, number of internal hydrogen bonds, and SASA calculated for HSA in the presence and absence of the Si$_3$N$_4$ monolayer in a contrasting fashion.
From Fig.~\ref{subfig:3(a)}, it is seen that the HSA structure has remained stable during the simulations with all-atom RMSDs of $<0.5$ and 0.6 nm from the reference crystal structure (3B9M) on interaction with the monolayer (HSA-ML-W) and in aqueous lonely (HSA-W), respectively. The RMSD of HSA-ML-W also takes smaller values compared to the other over the last 35 ns, demonstrating that interaction with Si$_3$N$_4$ nanosheet considerably reduces thermal fluctuations of HSA due to binding to the nanosheet, and accordingly decreases the protein's conformational change more than that of HSA-W.

Time dependence of the radius of gyration illustrated in Fig.~\ref{subfig:3(b)} also reveals the fact that the overall structural compactness of HSA on interaction with the monolayer is slightly (by $\sim 1.0$ \AA) smaller than those of HSA-W or the initial crystal structure. The per-residue RMSF [Fig.~\ref{subfig:3(c)}] averaged over all frames further indicates that interaction with the nanosheet considerably decreases the flexibility of all regions of HSA, and therefore makes its secondary structure resistant to any change raised by thermal fluctuations in aqueous. Consistently, the number of hydrogen bonds within the protein in both HSA-W and HSA-ML-W [Fig.~\ref{subfig:3(d)}] exhibit no remarkable change over the entire trajectory compared to each other. More precisely, the number of hydrogen bonds averaged over the last 45 ns is about 147 for HSA-W and 139 for HSA-ML-W, indicating a decrease as negligible as 5.5\% on interaction with the monolayer. The calculated SASA [Fig.~\ref{subfig:3(e)}] over the last 40 ns of the two trajectories also show convergence in a way that HSA-ML-W takes larger values compared to HSA-W on average, in agreement with the gyration radius [Fig.~\ref{subfig:3(b)}]. As a result, HSA strongly binds onto the surface of Si$_3$N$_4$ monolayer, forming a stable complex. 

\begin{figure}[H]
	\centering
	\fbox{\rule[0cm]{0cm}{0cm}     \rule[0cm]{0cm}{0cm}
	\subfigure[HSA-W]{
		\includegraphics[scale=0.49,angle=0]{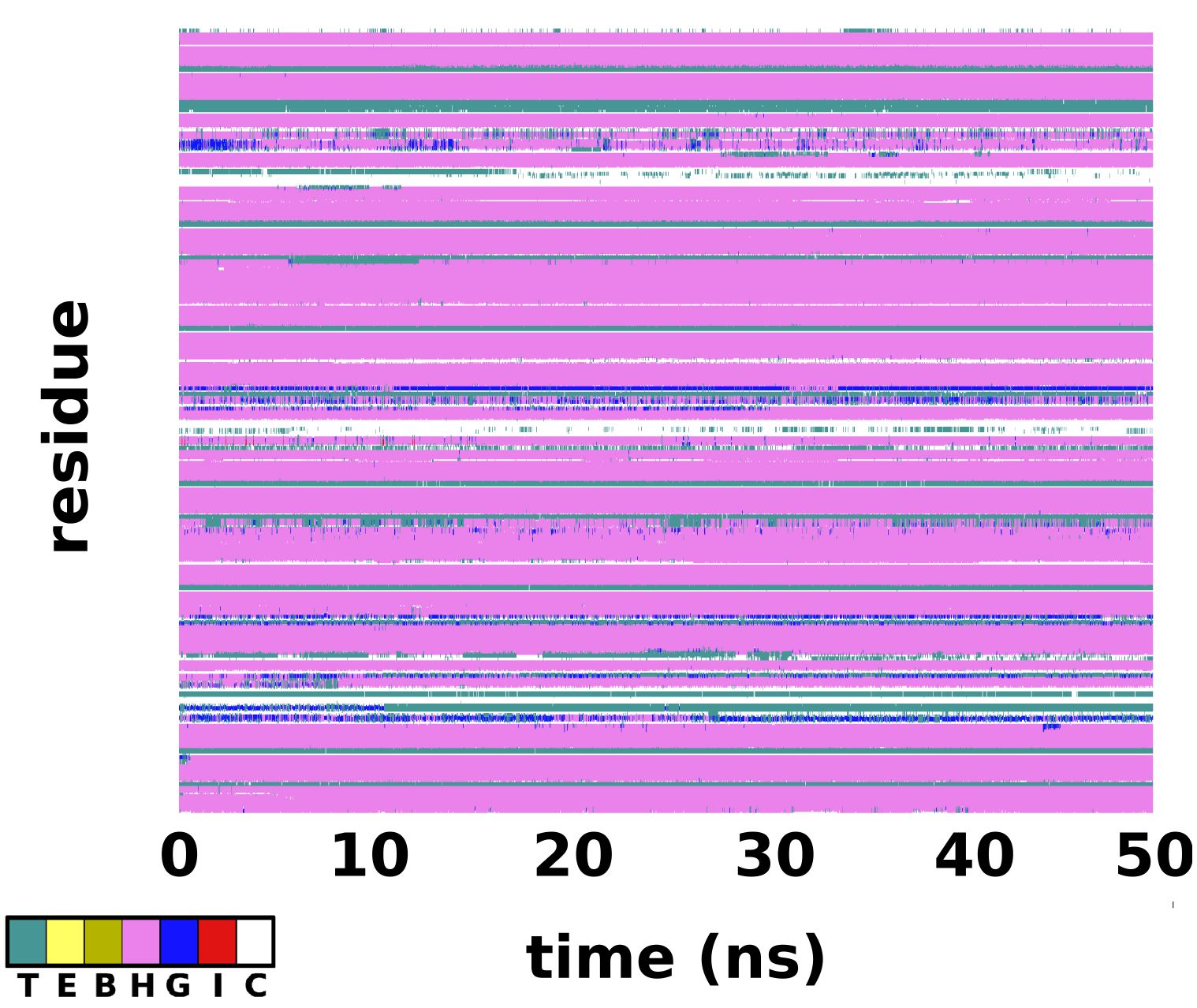}}
	\subfigure[HSA-ML-W]{
		\includegraphics[scale=0.495]{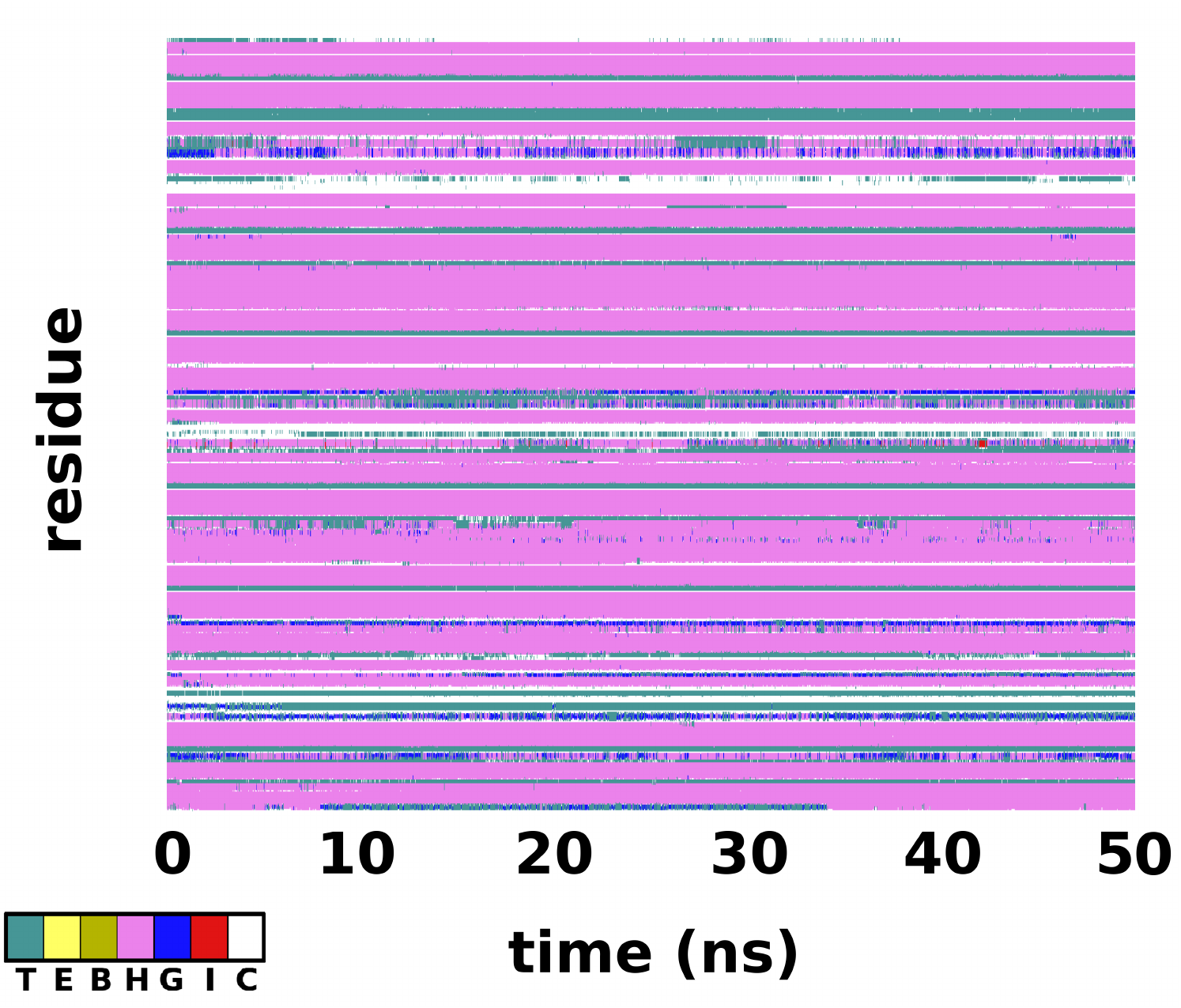}}}
	\caption{\label{fig:4}
		Variation in the secondary structure of HSA in (a) water (HSA-W), and (b) on interaction with the Si$_3$N$_4$ monolayer (HSA-ML-W) over the total time-spans. No significant change is observed in agreement with the previous findings. T, E, B, H, G, I, and C also stand for turn, extended configuration, isolated bridge, $\alpha$-helix, 3$_{10}$-helix, $\pi$-helix, and coil, respectively. The $\alpha$-helix-rich structure (the predominant, violet areas) of HSA is unmistakably clear.}
\end{figure}

The secondary structure of HSA as a function of time has also been provided in Fig.~\ref{fig:4} in the presence and absence of the silicon nitride monolayer.
Consistent with the previous analyses, no dramatic change is accordingly observed, and the $\alpha$-helix-rich structure of HSA [Fig.~\ref{subfig:1(a)}] is clearly preserved on interaction with the monolayer.

\begin{figure}[H]
	\centering
		\fbox{\rule[0cm]{0cm}{0cm}     \rule[0cm]{0cm}{0cm}
	\subfigure[HSA-W at $t=0$]{
		\includegraphics[scale=0.6,angle=0]{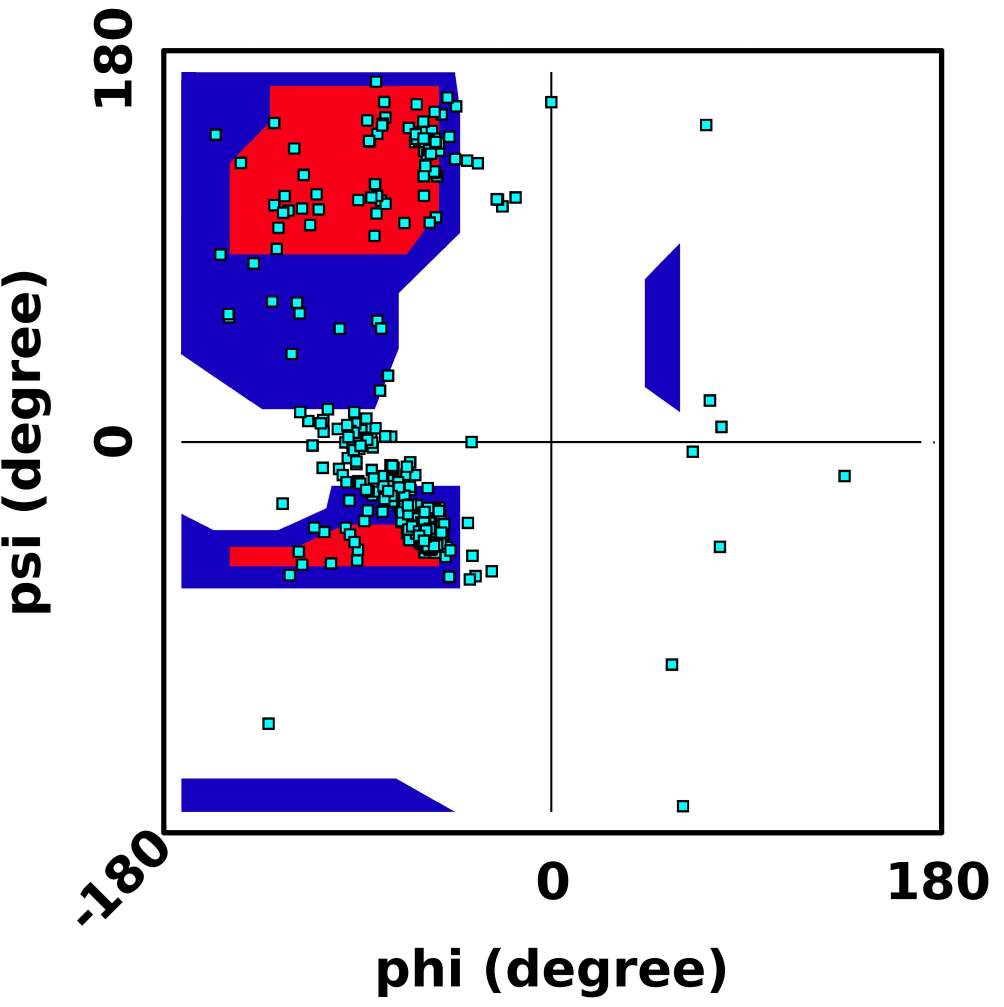}}
	\subfigure[HSA-W at $t=50$ ns]{
		\includegraphics[scale=0.6,angle=0]{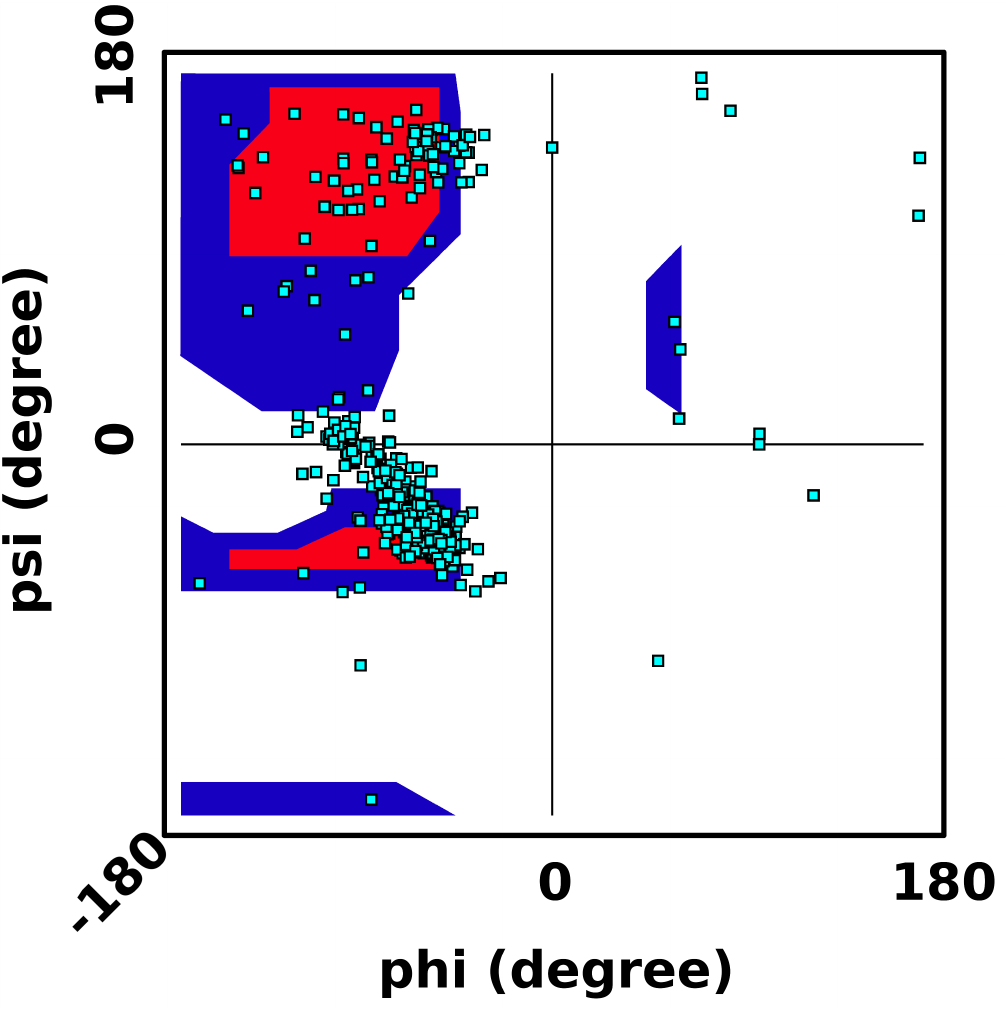}}}
		\fbox{\rule[0cm]{0cm}{0cm}     \rule[0cm]{0cm}{0cm}
	\subfigure[HSA-ML-W at $t=0$]{
		\includegraphics[scale=0.6,angle=0]{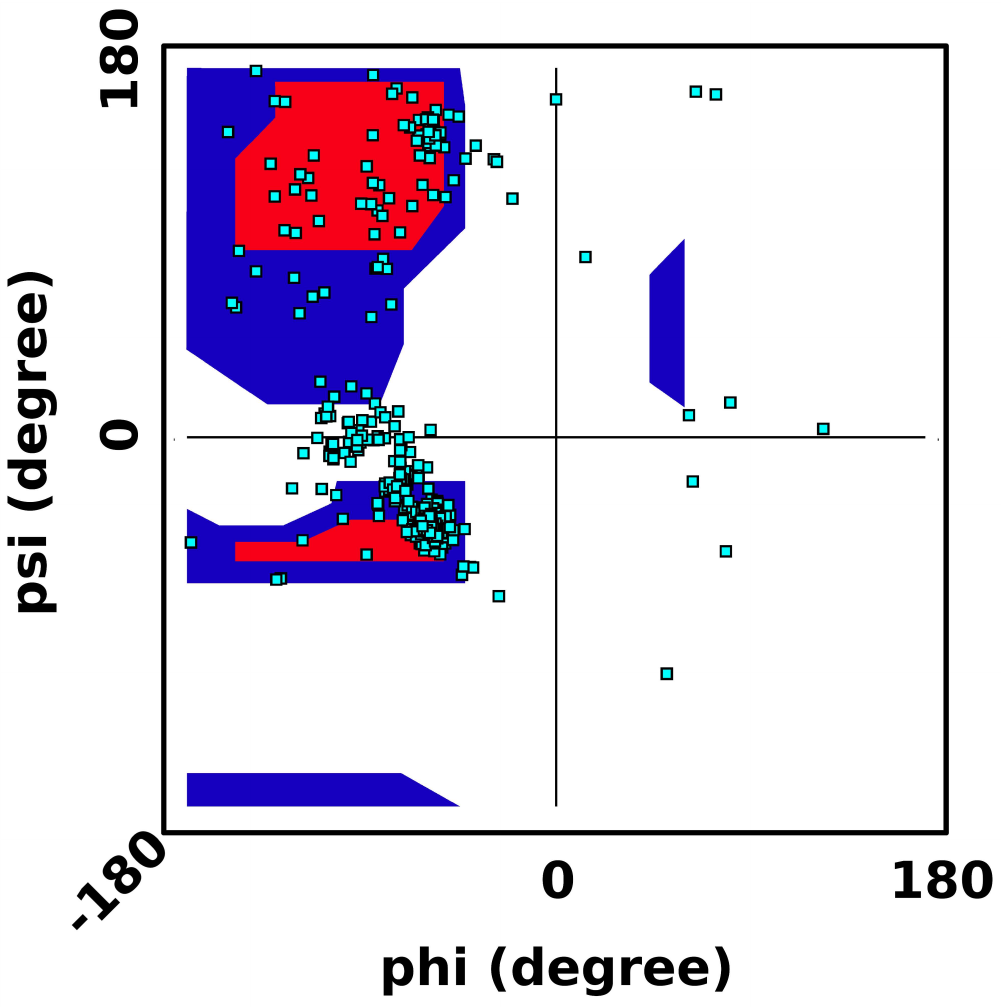}}
	\subfigure[HSA-ML-W at $t=50$ ns]{
		\includegraphics[scale=0.6,angle=0]{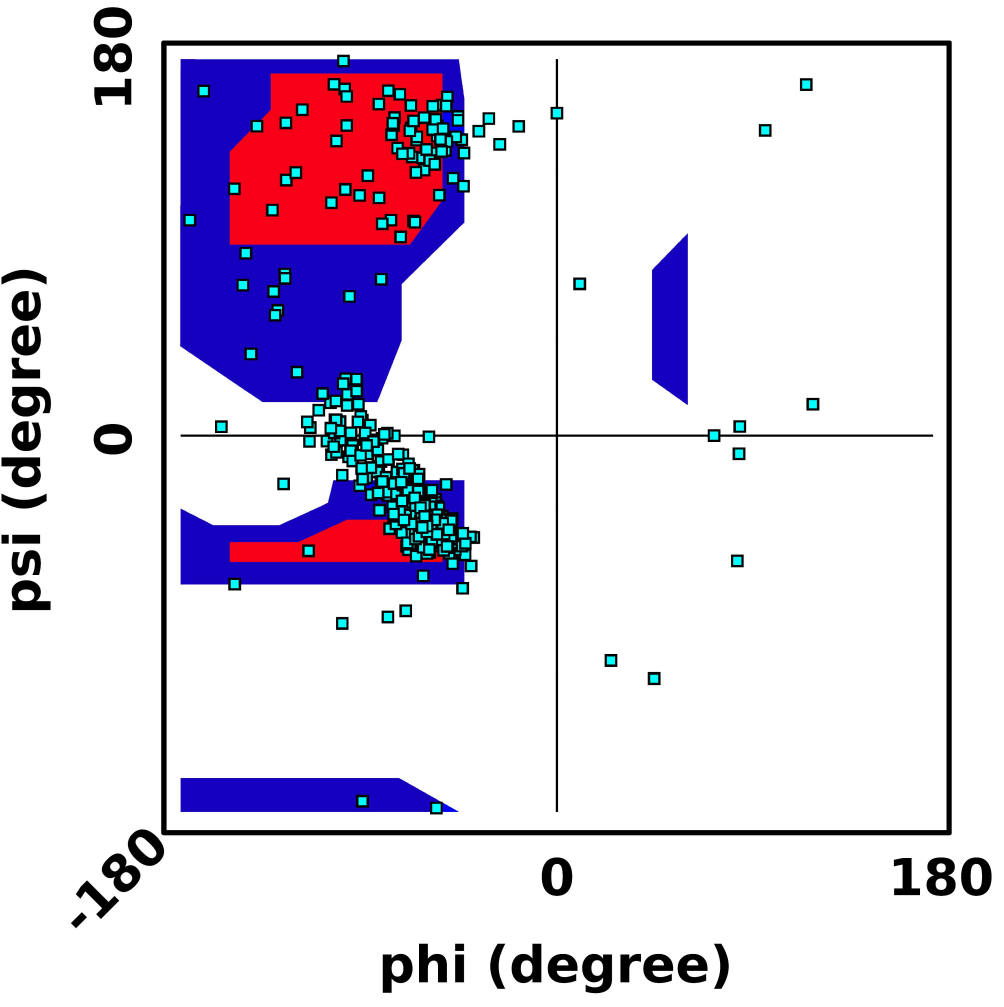}}}
	\caption{\label{fig:5}
		The Ramachandran plots obtained for HSA-W and HSA-ML-W at (a,\hspace{0.5mm}c) $t=0$ and (b,\hspace{0.5mm}d) $t=50$ ns, respectively. The white, red, and blue areas are respectively the sterically disallowed region for amino acids, the allowed regions corresponding to the helix and extended conformations, and the outer limit regions. The cyan points also indicate the distribution of dihedral angle values of each residue in the $\phi-\psi$ space.}
\end{figure}

Examining the corresponding Ramachandran plots also confirms the preceding observations as illustrated in Fig.~\ref{fig:5}.
As is seen, the distributions of dihedral angles (dominantly on the right-handed $\alpha$-helix region) over different areas are nearly the same for both HSA-W and HSA-ML-W at $t=0$ and 50 ns. The distributions associated to HSA-W and HSA-ML-W at $t=50$ ns are also nearly the same, indicating that the secondary structure of HSA remains intact on interaction with 2D Si$_3$N$_4$ nanostructure.

\begin{figure}[H]
	\centering
		\fbox{\rule[0cm]{0cm}{0cm}     \rule[0cm]{0cm}{0cm}
	\subfigure[]{\label{subfig:6(a)}
		\includegraphics[scale=0.28]{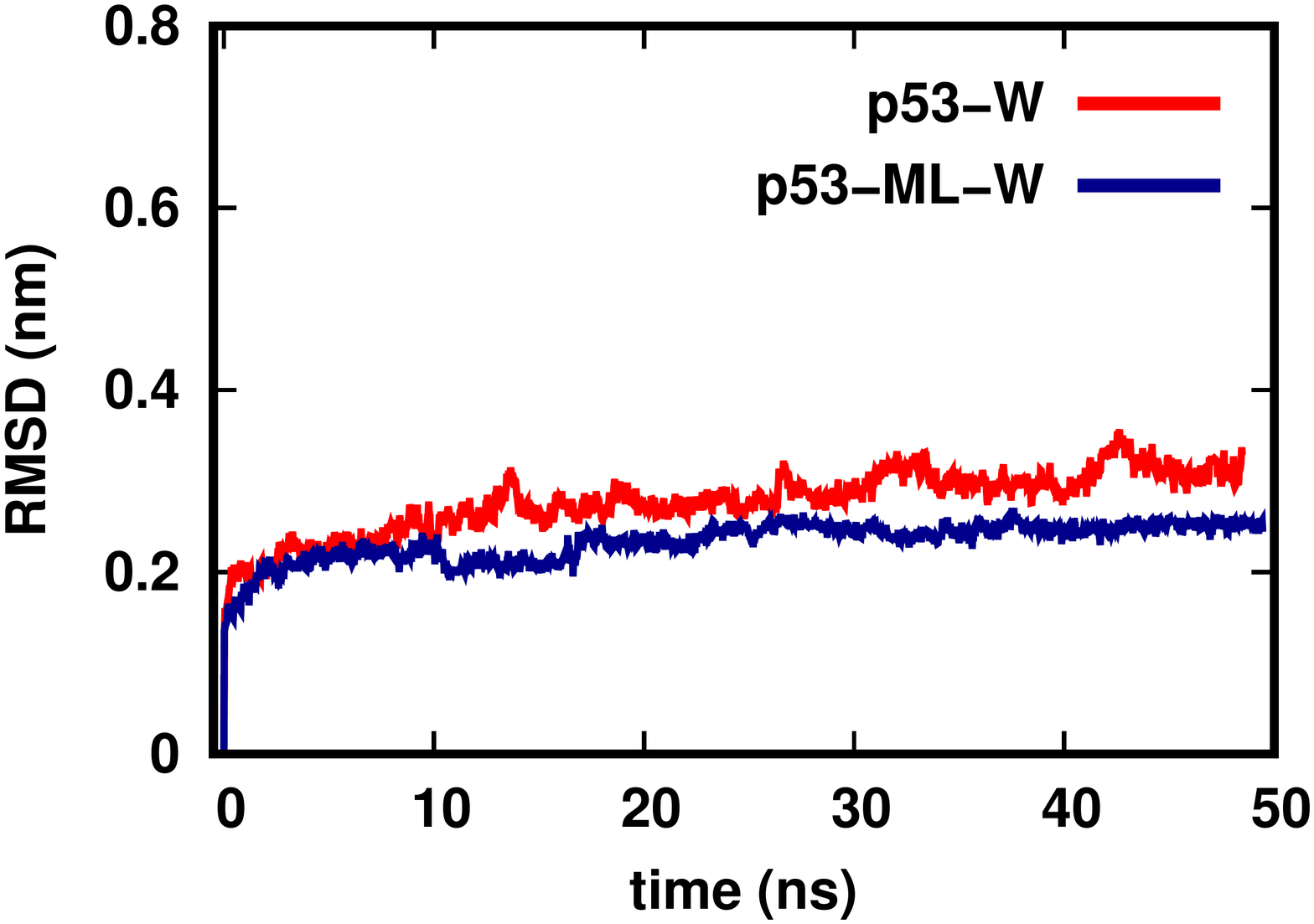}}
	\subfigure[]{\label{subfig:6(b)}
		\includegraphics[scale=0.28]{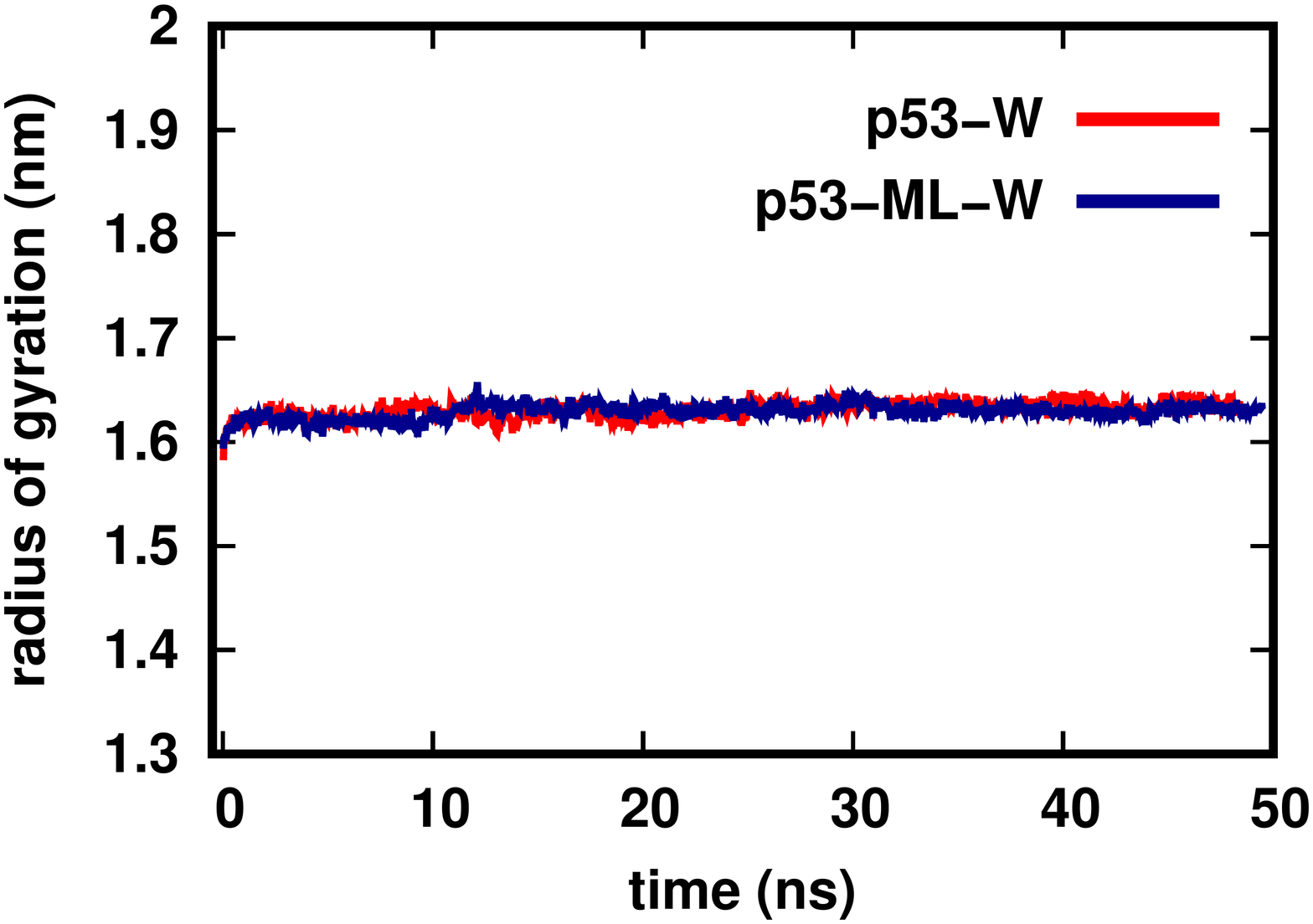}}}
	\fbox{\rule[0cm]{0cm}{0cm}     \rule[0cm]{0cm}{0cm}
	\subfigure[]{\label{subfig:6(c)}
		\includegraphics[scale=0.28]{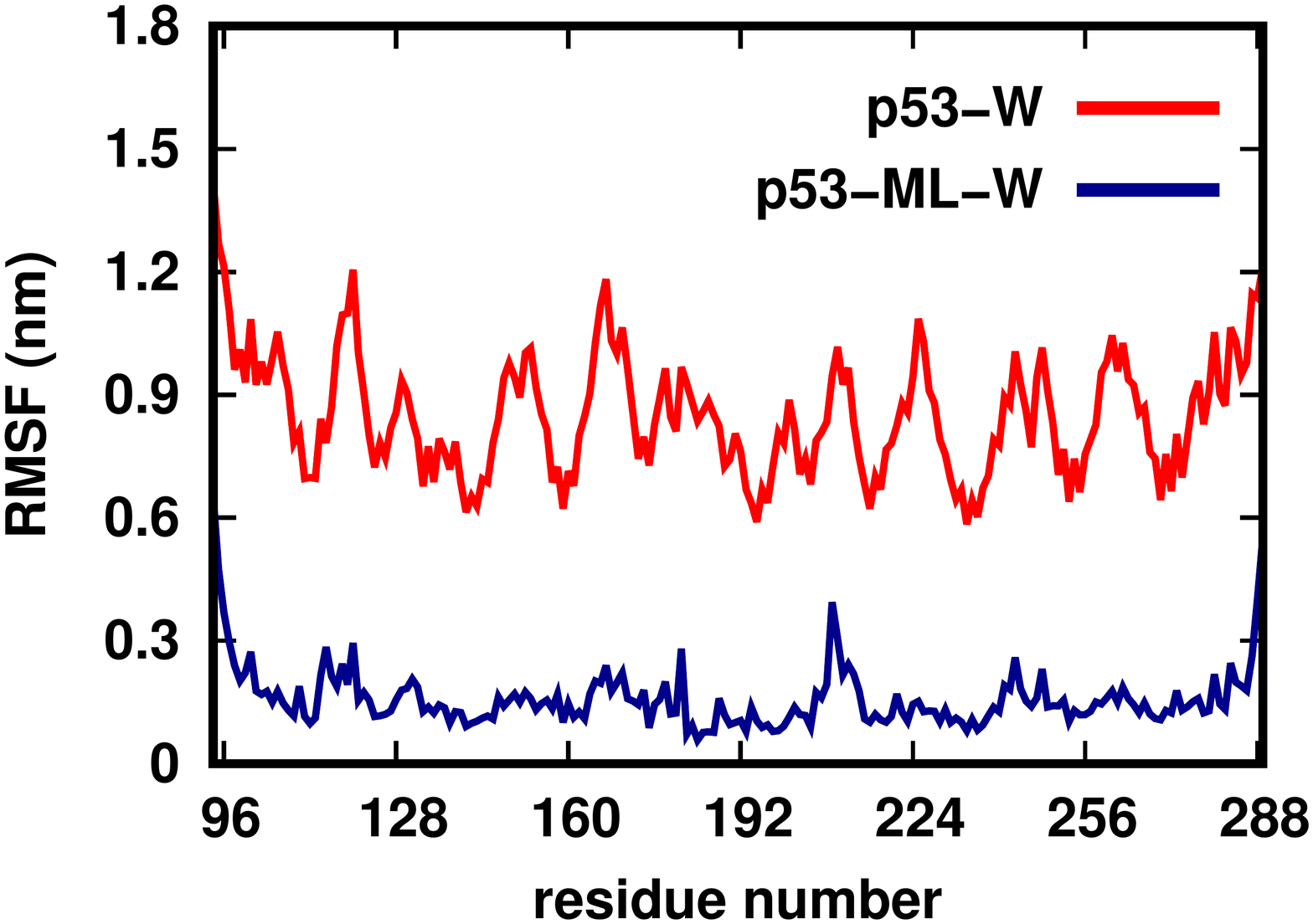}}
	\subfigure[]{\label{subfig:6(d)}
		\includegraphics[scale=0.28]{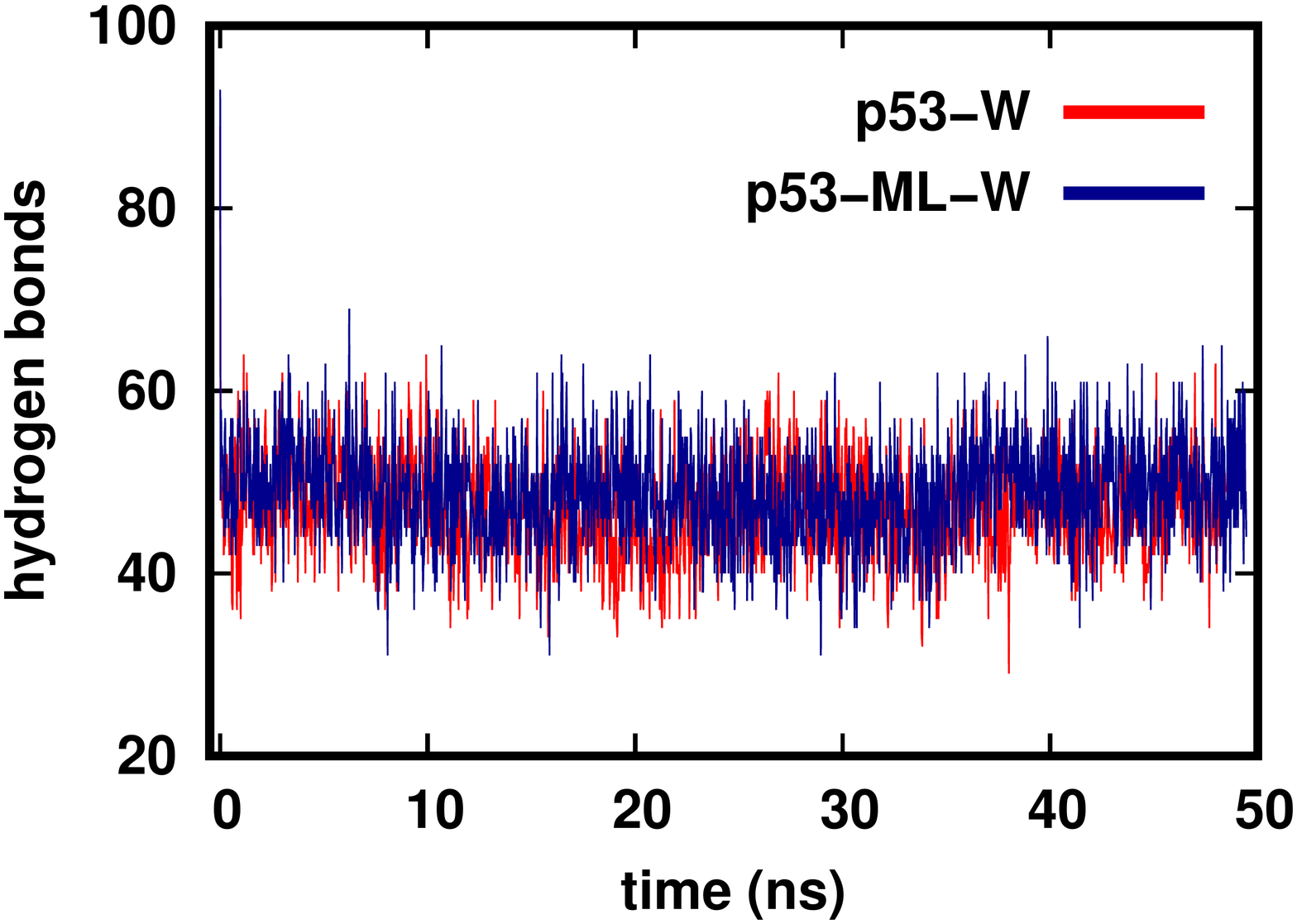}}}
\fbox{\rule[0cm]{0cm}{0cm}     \rule[0cm]{0cm}{0cm}
	\subfigure[]{\label{subfig:6(e)}
		\includegraphics[scale=0.28]{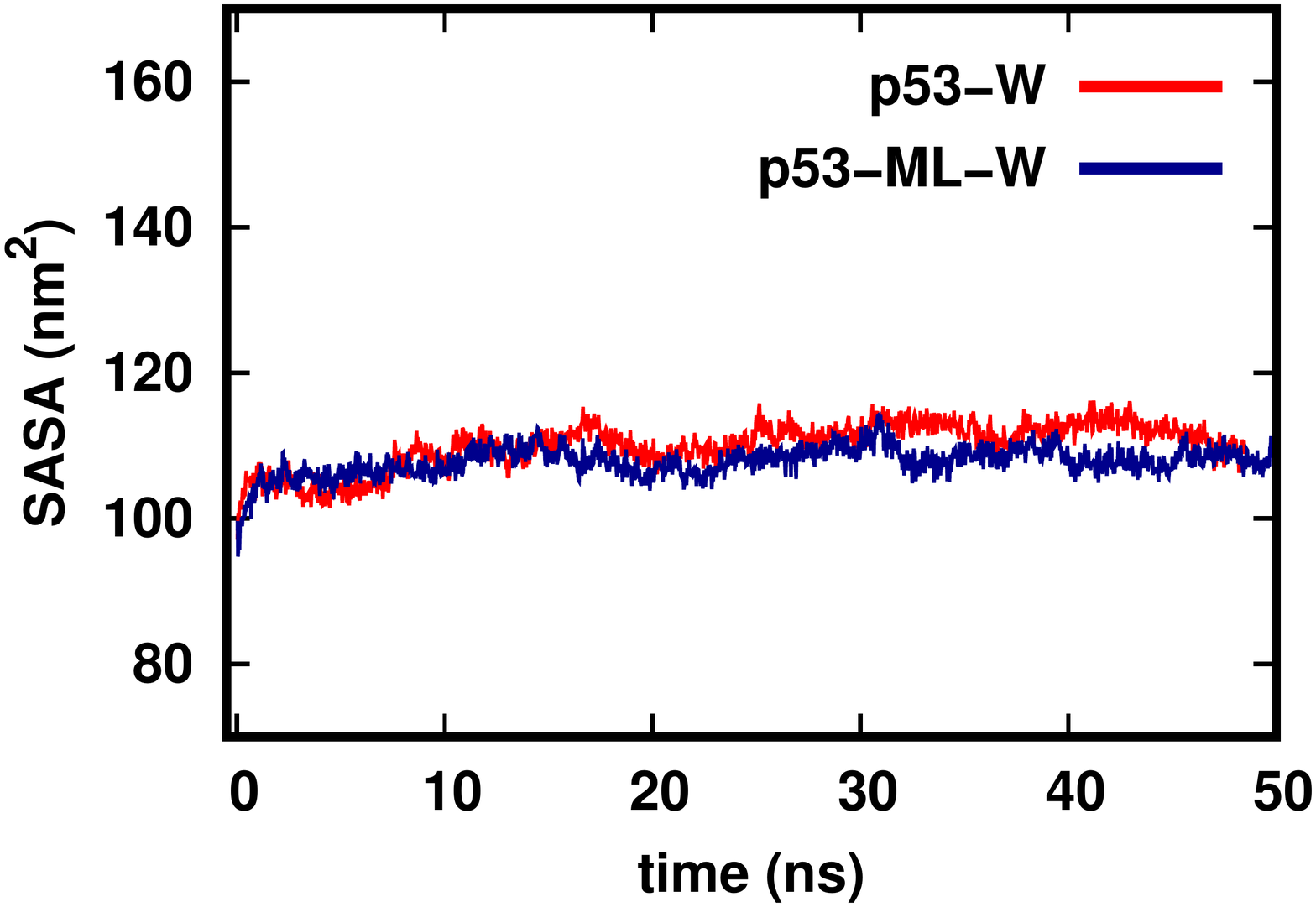}}}
	\caption{\label{fig:6}
		(a) RMSD, (b) radius of gyration, (c) per-residue RMSF, (d) number of hydrogen bonds, and (e) SASA obtained for p53 in the absence (p53-W, red) and presence (p53-ML-W, blue) of the monolayer.}
\end{figure}

Similar findings were obtained for p53. Comparing the associated all-atom RMSD curves in the absence and presence of the silicon nitride nanosheet [Fig.~\ref{subfig:6(a)}] indicates that the related secondary structure also remains unchanged during the last 40 ns with values of about 0.25 and 0.3 nm for p53-ML-W and p53-W, respectively.
That the RMSD of p53-ML-W takes smaller values over the entire time-span reveals that interaction with 2D Si$_3$N$_4$ dramatically decreases fluctuations of the protein caused by thermal energy, therefore, abates the tendency toward any conformational change. 

Examining the time dependence of radius of gyration [Fig.~\ref{subfig:6(b)}] further indicates that the overall structural compactness of p53 in the presence and absence of the monolayer are nearly the same over the entire trajectory. The RMSF curves, illustrated in Fig.~\ref{subfig:6(c)}, show a considerable decrease in the fluctuating pattern values by $\sim 0.7$ nm, and therefore in the overall flexibility of the protein on interaction with the monolayer. The peak in the middle (at residue 209) of the p53-ML-W curve is for the arginine residue, which exhibits a high degree of flexibility at this point based on the fact that it is located on a turn secondary structure with a distance of about 2.8 nm from the monolayer, indicating no effective binding between them as well.

Interaction with Si$_3$N$_4$ nanosheet has also no (significant) impact on the number of hydrogen bonds within p53, and consequently on the secondary structure of p53 as seen in Fig.~\ref{subfig:6(d)}---the average values of the number of hydrogen bonds over the last 45 ns are 47 for p53-W and 49 for p53-ML-W, indicating an increase as small as 3.9\% on interaction with the monolayer.

The SASA curves [Fig.~\ref{subfig:6(e)}] in the presence and absence of the monolayer show nearly the same converging patterns over the last 40 ns of the trajectories. Nonetheless, a negligible difference between the two curves could be observed from 18 ns on, with smaller values for that of p53-ML-W consistent with the related RMSD [Fig.~\ref{subfig:6(a)}] and gyration radius [Fig.~\ref{subfig:6(b)}]. Indeed, such a discrepancy is not so important as to considerably change the secondary structure of the protein.

\begin{figure}[H]
	\centering
\fbox{\rule[0cm]{0cm}{0cm}     \rule[0cm]{0cm}{0cm}
	\subfigure[p53-W]{
		\includegraphics[scale=0.49]{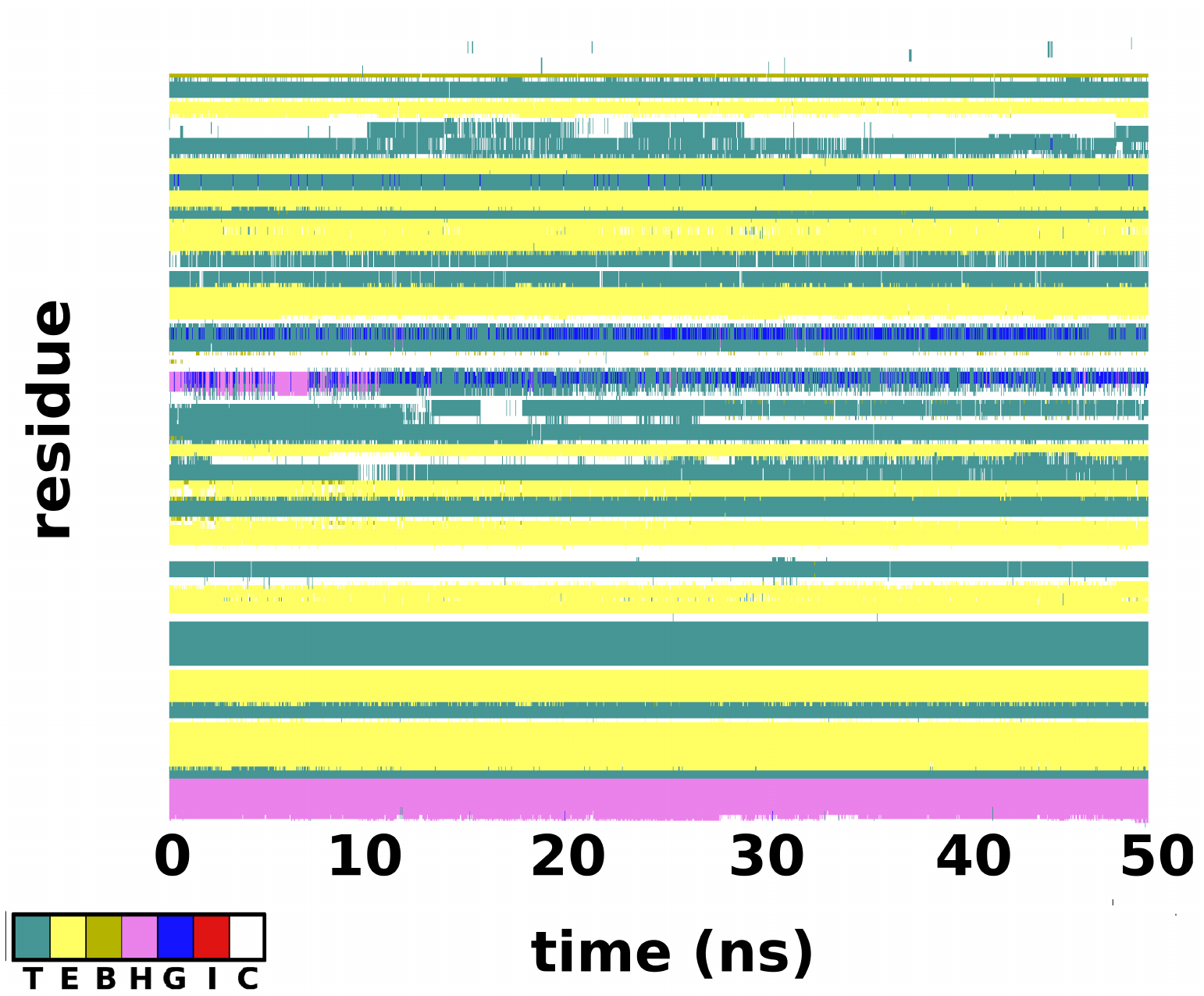}}
	\subfigure[p53-ML-W]{
		\includegraphics[scale=0.49]{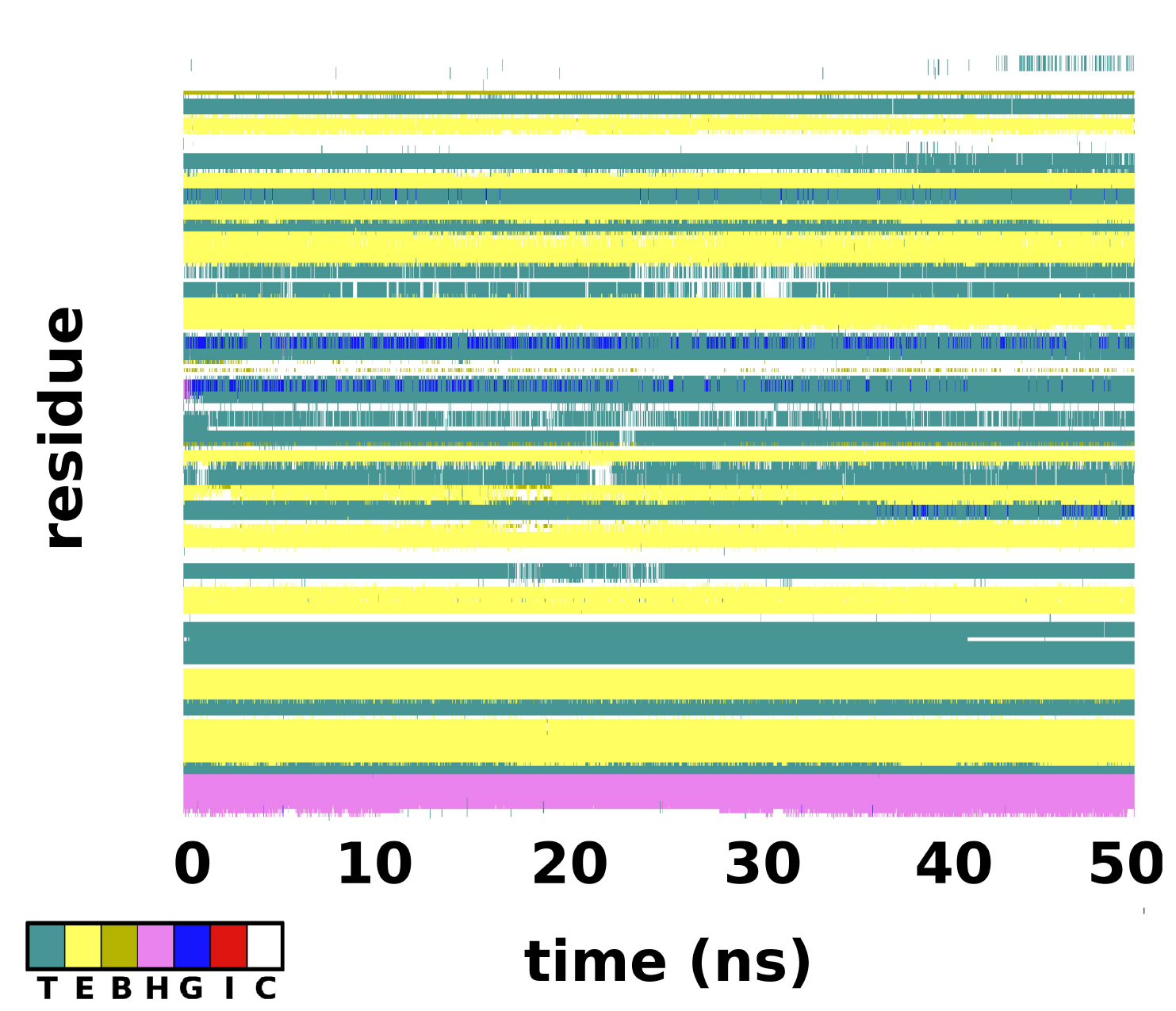}}}
	\caption{\label{fig:7}
		Secondary structure of p53 in (a) water (p53-W), and (b) on interaction with the Si$_3$N$_4$ monolayer (p53-ML-W), showing no significant discrepancy, consistent with the previous analysis. All the secondary structure types particularly the $\alpha$-helix, the extended configuration, and the turn, have remained relatively intact in the absence and presence of the nanosheet.}
\end{figure}

The secondary structure of p53 has been illustrated in Fig.~\ref{fig:7} in the presence and absence of the silicon nitride monolayer over the entire trajectory.
Nearly the same, $\beta$-rich patterns is observed in the two subfigures consistent with the tertiary structure of p53 [Fig.~\ref{subfig:1(b)}], demonstrating the minimal, insignificant change in the related secondary structure on interaction with the monolayer.

\begin{figure}[H]
	\centering
	\fbox{\rule[0cm]{0cm}{0cm}     \rule[0cm]{0cm}{0cm}
	\subfigure[p53-W at $t=0$]{
		\includegraphics[scale=0.6,angle=0]{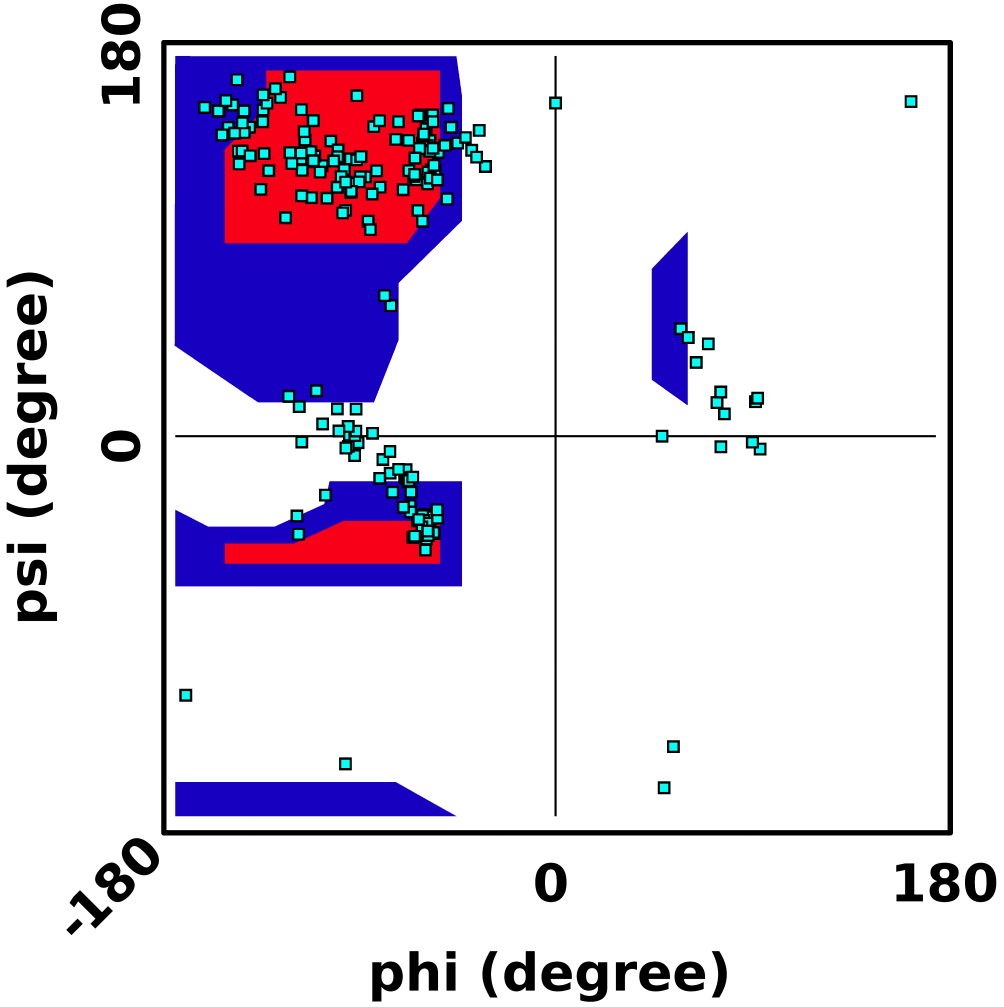}}
	\subfigure[p53-W at $t=30$ ns]{
		\includegraphics[scale=0.6,angle=0]{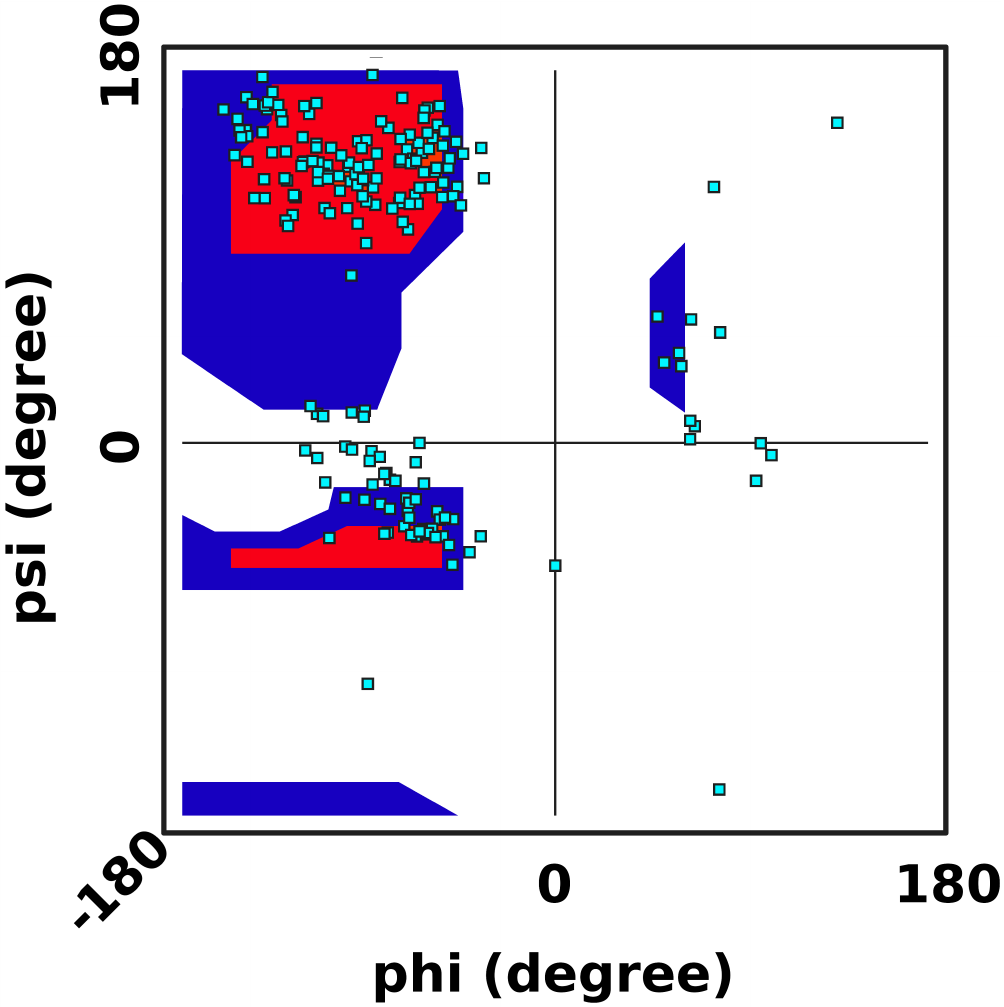}}}
		\fbox{\rule[0cm]{0cm}{0cm}     \rule[0cm]{0cm}{0cm}
	\subfigure[p53-ML-W at $t=0$]{
		\includegraphics[scale=0.6,angle=0]{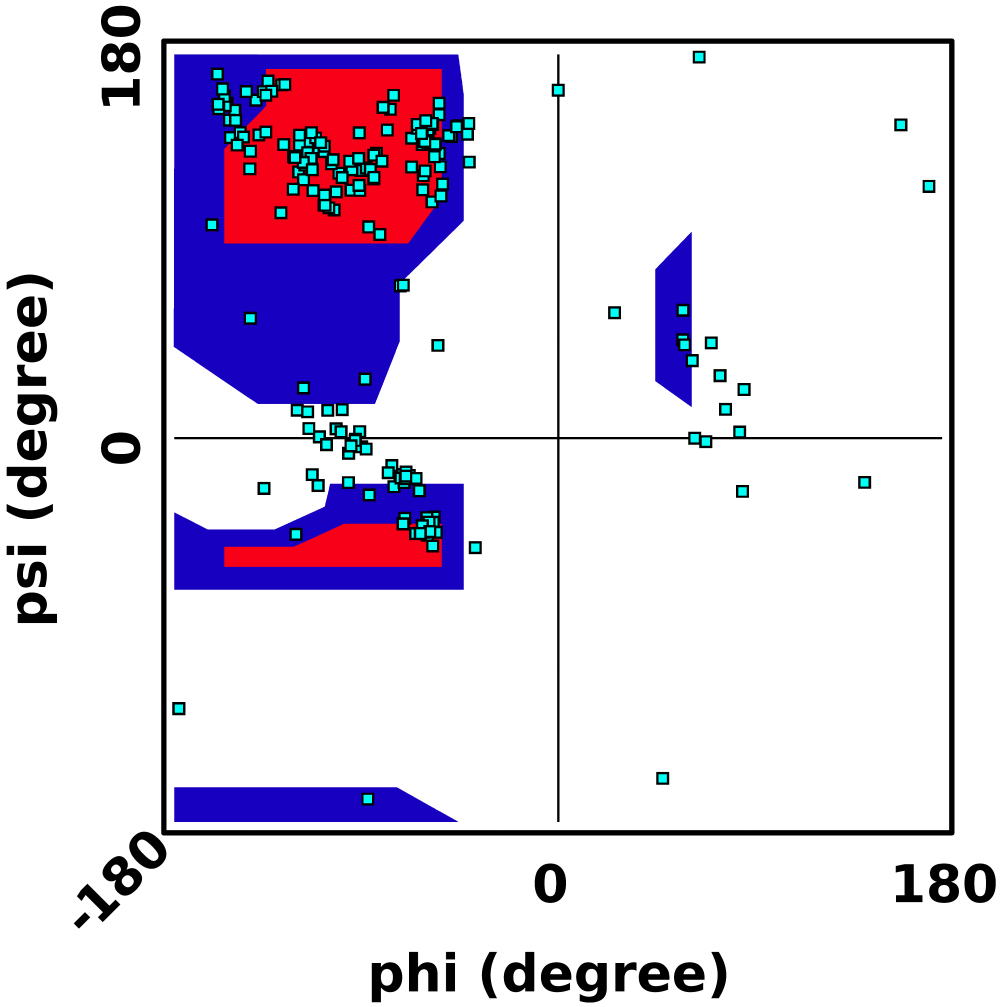}}
	\subfigure[p53-ML-W at $t=30$ ns]{
		\includegraphics[scale=0.6,angle=0]{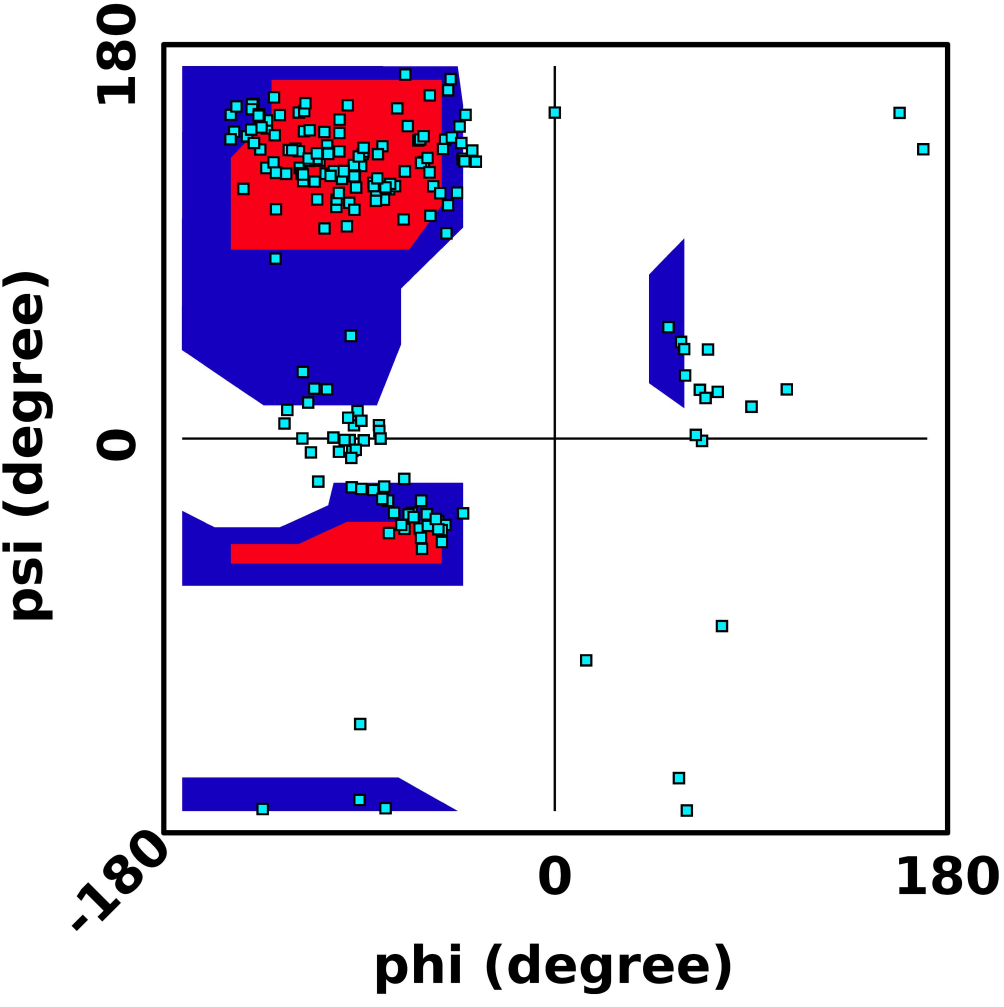}}}
	\caption{\label{fig:8}
		The Ramachandran plots for p53-W and p53-ML-W at (a,\hspace{0.5mm}c) $t=0$ and (b,\hspace{0.5mm}d) $t=50$ ns.}
\end{figure}

The associated Ramachandran plots illustrated in Fig.~\ref{fig:8} are also in agreement with the previous findings.
In the same way as HSA, but to a higher extent, the distributions of the dihedral angles (dominantly on the extended region) over different areas are very close to each other comparing p53-W and p53-ML-W at $t=0$ and 50 ns, and comparing p53-W and p53-ML-W at $t=50$ ns. As a result, no considerable change in the related secondary structure could be observed in case it is in contact with the monolayer.

\begin{figure}[H]
	\centering
		\fbox{\rule[0cm]{0cm}{0cm}     \rule[0cm]{0cm}{0cm}
	\includegraphics[scale=0.4,angle=-90]{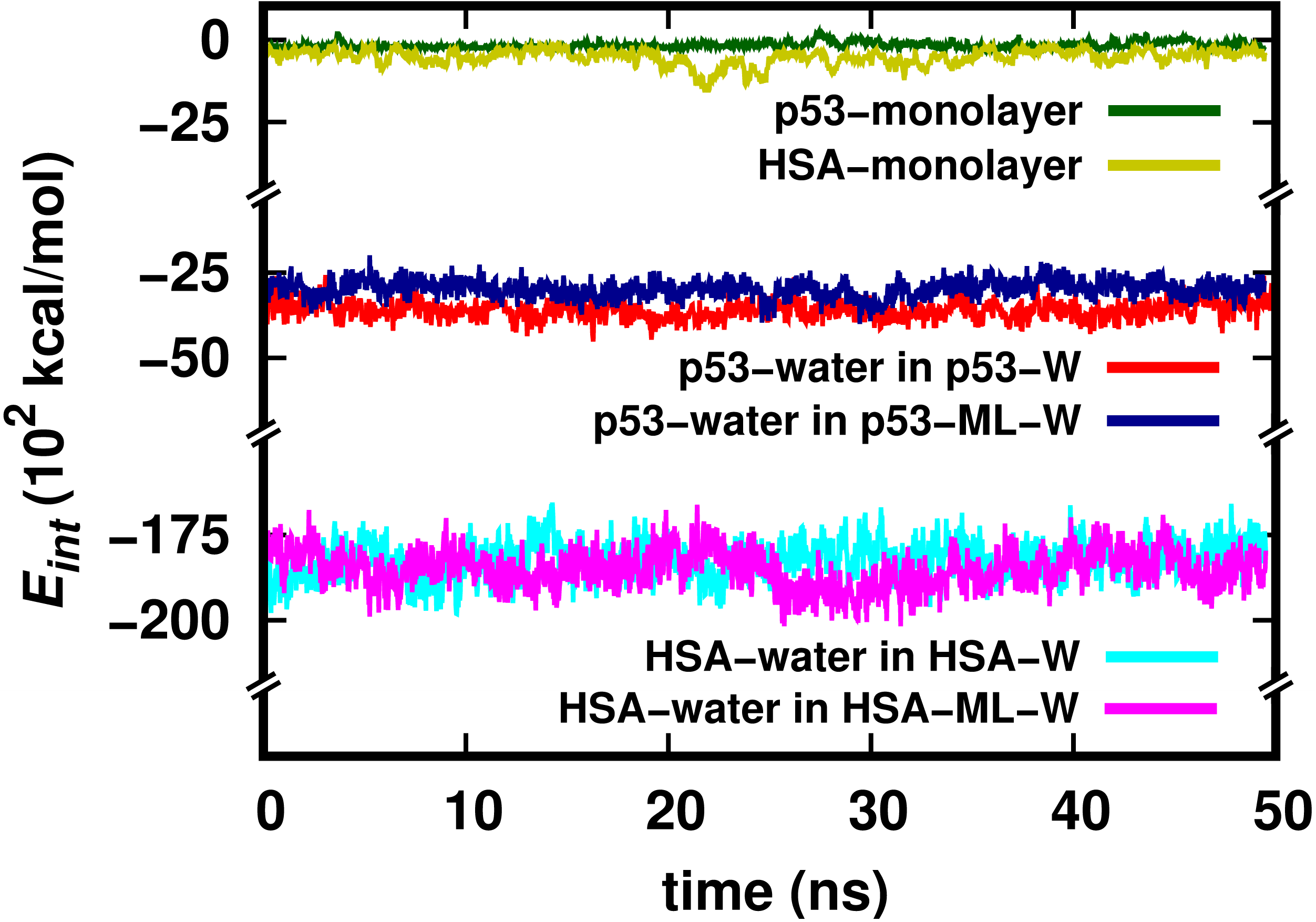}}
	\caption{\label{fig:9}
		The interaction energies including protein-monolayer as well as protein-water in the presence and absence of the Si$_3$N$_4$ monolayer as functions of time over the total time-span.}
\end{figure}

We finally estimated the time dependence of the interaction energy ($E_{int}$) between protein-monolayer on one hand, and protein-water in the presence and absence of the monolayer on the other, as illustrated in Fig.~\ref{fig:9}.
The $E_{int}$ values for p53- and HSA-monolayer are respectively about $-151$ and $-510$ kcal$\big/$mol, which are dramatically larger (therefore weaker bindings) than those of the protein-water interactions, showing that these biological proteins do not make stable complexes on interaction with the monolayer. That HSA has the lower value is also an indication of the fact that it is a considerably larger protein compared to p53, which accordingly led to larger SASA values comparing Figs.~\ref{subfig:3(e)} and~\ref{subfig:6(e)}.

Interaction with the Si$_3$N$_4$ nanosheet also decreases the p53-water binding by about 16.7\% from $-3574$ (in p53-W) to $-2977.7$ kcal$\big/$mol (in p53-ML-W). In contrast, the presence of the monolayer increases the HSA-water binding by 1.85\% from $-18201.4$ (in HSA-W) to $-18537.7$ (in HSA-ML-W). However, non of these percent values is significant so as to dramatically affect the protein-water bindings.
\section{\label{sec:5}Conclusions}
All-atom molecular dynamics simulations were applied to investigate the biocompatibility of 2D, hexagonal $\beta$-Si$_3$N$_4$ monolayer via examining its possible impacts on both HSA (human serum albumin) and p53 antitumor protein. We accordingly calculated and examined a number of important MD indicators including RMSD, radius of gyration, per-residue RMSF, number of hydrogen bonds, solvent-accessible surface area, and secondary structure of each protein in a contrasting fashion in the presence and absence of the monolayer. Results verified that the secondary structures of these proteins remain nearly intact on interaction with Si$_3$N$_4$ nanosheet. Examining the associated protein-monolayer and protein-water interaction energies in the presence and absence of the monolayer further revealed that these biological proteins do not make stable complexes with the monolayer. The presence of Si$_3$N$_4$ also affected both HSA- and p53-water bindings very marginally. It was accordingly inferred that hexagonal $\beta$-Si$_3$N$_4$ nanosheet is indeed a biocompatible material and could then be used as a therapeutic or carrier for in vivo applications.
\section*{Declaration of Competing Interest}
The authors declare that they have no known competing financial interests or personal relationships that could have appeared to influence the work reported in this paper.
\section*{Data Availability Statement}
The data supporting the findings of the present investigation are available on reasonable request from the corresponding author.

\end{document}